\documentclass[a4paper,11pt]{article}
\pdfoutput=1 %

\usepackage{jheppub} %
\usepackage[utf8x]{inputenc}
\usepackage{mathrsfs}
\usepackage{amssymb}
\usepackage{amsmath}
\usepackage{mathtools}
\usepackage{slashed}
\usepackage{graphicx}
\usepackage{grffile}
\usepackage{physics}

\usepackage{subcaption}
\usepackage[normalem]{ulem}
\usepackage{xcolor}
\definecolor{lcolor}{rgb}{0.5,0,0}
\definecolor{citcolor}{rgb}{0,0.3,0.0}

\usepackage{tikz}

\newcommand{\rt}{{\mathbf{r}}}
\newcommand{\xt}{{\mathbf{x}}}
\newcommand{\Pt}{{\mathbf{P}}}
\newcommand{\bt}{{\mathbf{b}}}

\newcommand{\nc}{N_\mathrm{c}}

\newcommand{\xpom}{{x_\mathbb{P}}}
\newcommand{\zmin}{z_{\text{min}}}
\newcommand{\Ydip}{Y_\text{dip}}
\newcommand{\Yqqg}{Y_{\text{q}\bar{\text{q}}\text{g}}}
\newcommand{\nlodip}{\mathrm{NLO}_\text{dip}}
\newcommand{\nloqqg}{\mathrm{NLO}_{\text{q}\bar{\text{q}}\text{g}}}
\newcommand{\lo}{\mathrm{LO}}
\newcommand{\nlo}{\mathrm{NLO}}

\newcommand{\Ybk}{{Y_{0,\text{BK}}}}
\newcommand{\etabk}{{\eta_{0,\text{BK}}}}
\newcommand{\OLDME}{\langle \mathcal{O}_1\rangle_V}
\newcommand{\qLDME}{\langle \vec q^{\,2} \rangle_V}

\newcommand{\B}{/(\xt_{01}^2 e^{\gamma_E})}

\newcommand{\gev}{\ \textrm{GeV}}

\newcommand{\qs}{Q_\mathrm{s}}

\newcommand{\xbj}{{x_\text{bj}}}

\newcommand{\acal}{\mathcal{A}}
\newcommand{\fcal}{\mathcal{F}}
\newcommand{\ncal}{\mathcal{N}}
\newcommand{\vcal}{\mathcal{V}}
\newcommand{\ical}{\mathcal{I}}
\newcommand{\jcal}{\mathcal{J}}
\newcommand{\scal}{\mathcal{S}}
\newcommand{\kcal}{\mathcal{K}}
\newcommand{\ocal}{\mathcal{O}}

\newcommand{\mcal}{\mathcal{M}}

\newcommand{\as}{\alpha_\mathrm{s}}
\newcommand{\der}{\mathrm{d}}

\newcommand{\Deltat}{\mathbf{\Delta}}

\newcommand{\jpsi}{$\mathrm{J}/\psi$ }
\newcommand{\jpsim}{\mathrm{J}/\psi}

\begin{document}

\author[a,b]{Heikki Mäntysaari}
\author[a,b]{Jani Penttala}

\affiliation[a]{
Department of Physics, University of Jyväskylä %
 P.O. Box 35, 40014 University of Jyväskylä, Finland
}
\affiliation[b]{
Helsinki Institute of Physics, P.O. Box 64, 00014 University of Helsinki, Finland
}

\emailAdd{heikki.mantysaari@jyu.fi}
\emailAdd{jani.j.penttala@jyu.fi}

\title{

Complete calculation of exclusive heavy vector meson production at next-to-leading order in the dipole picture
}

\preprint{}

\abstract{
Exclusive production of transversely polarized heavy vector mesons in deep inelastic scattering at high energy is calculated at next-to-leading order accuracy in the Color Glass Condensate framework. In addition to the first QCD correction proportional to the strong coupling constant $\as$, we systematically also include the first relativistic correction proportional to the heavy quark velocity squared $v^2$. When combined with our previously published results for longitudinal vector meson production at next-to-leading order accuracy, these results make phenomenological calculations of heavy vector meson production possible at the order $\mathcal{O}(\as v^0, \as^0 v^2)$. When applied to \jpsi and $\Upsilon$ production at HERA and at the LHC, a good agreement between the next-to-leading order calculations and experimental data is found. Additionally, we demonstrate that vector meson production can provide additional constraints compared to structure function analyses when the nonperturbative initial condition for the Balitsky-Kovchegov evolution equation is extracted.

}

\maketitle

\section{Introduction}
\label{sec:introduction}

In Quantum Chromodynamics (QCD), emission of small momentum fraction $x$ gluons is preferred, which renders parton densities very large when probed at small momentum fraction $x$ in high-energy collider experiments~\cite{Aaron:2009aa,Abramowicz:2015mha}. Consequently, parton densities eventually become so large that the smallness of the QCD coupling $\as$ is compensated by the gluon density, and non-linear dynamics starts to dominate in the hadron wave function~\cite{Morreale:2021pnn}.

An especially powerful probe of non-linear QCD dynamics is given by exclusive vector meson production in deep inelastic scattering (DIS) experiments. The exclusivity of the process requires that, at leading order, two gluons are exchanged with the target hadron at the amplitude level. Thus the cross section approximately scales as gluon density squared~\cite{Ryskin:1992ui} (at next-to-leading order in collinear factorization the relationship is less direct, see Ref.~\cite{Eskola:2022vpi}) An additional advantage is that only in exclusive scattering processes it is possible to determine the total transverse momentum transfer $\Deltat$, which is a Fourier conjugate to the impact parameter, and as such the spectra are sensitive to the spatial distribution of color charge in the target color field~\cite{Klein:2019qfb}.

Exclusive production of heavy vector mesons, in particular $\mathrm{J}/\psi$, has been studied in detail in electron-proton DIS experiments at HERA~\cite{ZEUS:2004yeh,ZEUS:2002wfj,H1:2005dtp,Chekanov:2009ab,Breitweg:1999jy,Chekanov:2002rm,Aktas:2003zi,H1:2013okq}. Recently, it has also become possible to access even higher center-of-mass energies and scattering off nuclear targets in  ultra-peripheral collisions (UPCs) at RHIC~\cite{Adam:2019rxb,Afanasiev:2009hy,STAR:2021wwq} and at the LHC~\cite{Acharya:2021bnz,ALICE:2014eof,ALICE:2018oyo,Aaij:2013jxj,LHCb:2014acg,LHCb:2018rcm,ALICE:2019tqa,ALICE:2013wjo,ALICE:2012yye,ALICE:2021gpt,CMS:2016itn,LHCb:2021bfl}. 
In these events the impact parameter is larger than the sum of the radii of the colliding nuclei, which suppresses strong interactions and these events are effectively real photon-nucleus scattering processes~\cite{Bertulani:2005ru}.

In the next decade, the Electron-Ion Collider in the US~\cite{Aschenauer:2017jsk,AbdulKhalek:2021gbh} and other potential nuclear DIS facilities~\cite{LHeC:2020van,Anderle:2021wcy} will provide vast amounts of precise data on exclusive vector meson production over a wide kinematical domain. To take advantage of these recent and future developments that provide a unique access to non-linear QCD dynamics at small $x$, it is important to develop theoretical calculations to the comparable level of accuracy.

To describe QCD in the high energy (and density) regime, we employ the Color Glass Condensate (CGC) effective field theory approach~\cite{Iancu:2003xm,Gelis:2010nm}. In this formulation the color field of the target is written in terms of Wilson lines that describe an eikonal propagation of partons in the color field, resumming multiple interactions.
The purpose of this work is to present the first next-to-leading order (NLO) calculation of transversely polarized exclusive heavy vector meson production cross section at high energy within the CGC framework. 

Exclusive vector meson production has been studied extensively within the CGC framework at leading order in the QCD coupling, see for example Refs.~\cite{Kowalski:2006hc,Armesto:2014sma,Mantysaari:2018zdd,Mantysaari:2016jaz,Mantysaari:2016ykx} related to \jpsi production in $\gamma^*+p$ scattering and \cite{Lappi:2013am,Lappi:2010dd,Toll:2012mb,Caldwell:2010zza,Sambasivam:2019gdd,Mantysaari:2017slo,Mantysaari:2017dwh,Goncalves:2005yr,Bendova:2020hbb} in $\gamma^*+A$ scattering (e.g. in UPCs where the photon is real). Note also that in these leading-order calculations the small-$x$ evolution equations such as the Balitsky-Kovchegov (BK) equation~\cite{Balitsky:1995ub,Kovchegov:1999yj} (or phenomenological parametrizations modeling the small-$x$ evolution) resum contributions $\sim \as \ln 1/x$ to all orders, and running coupling corrections~\cite{Balitsky:2006wa,Kovchegov:2006vj} also resum a subset of higher-order contributions. 

At high energy the scattering process is conveniently described in the dipole picture  where the (virtual) photon splits into a quark-antiquark dipole long before the interaction with the target (see discussion in Sec.~\ref{sec:exclusive}). The dipole then interacts with the target and eventually forms the bound state. In order to develop the CGC calculations to NLO accuracy, all ingredients (the virtual photon and heavy vector meson wave functions, and the dipole-target scattering amplitude) are needed at this order in perturbation theory. In recent years there has been a rapid progress in the field to achieve this. The evolution equations at NLO, describing the center-of-mass energy dependence of the dipole-target scattering amplitude, are derived and solved in Refs.~\cite{Balitsky:2008zza,Balitsky:2013fea,Kovner:2013ona,Lappi:2020srm,Lappi:2016fmu,Lappi:2015fma} (and a subset of higher-order corrections are resummed in~\cite{Ducloue:2019ezk,Ducloue:2019jmy,Iancu:2015vea,Iancu:2015joa,Hatta:2016ujq}). The initial condition for the perturbative evolution is fitted to HERA structure function data at NLO accuracy in Ref.~\cite{Beuf:2020dxl} (see also Refs.~\cite{Dumitru:2021tvw,Dumitru:2020gla} for an NLO calculation of the proton color charge correlations at moderate $x$ that can potentially be used to initialize the evolution). The NLO light-front wave function for a virtual photon was first derived in the massless quark limit in Refs.~\cite{Beuf:2017bpd,Ducloue:2017ftk,Beuf:2016wdz,Balitsky:2010ze} and recently the results with finite quark masses have also become available~\cite{Beuf:2021qqa,Beuf:2021srj,Beuf:2022ndu}. The NLO wave functions exist also for heavy~\cite{Escobedo:2019bxn} and light~\cite{Boussarie:2016bkq,Mantysaari:2022bsp} vector mesons. In addition to structure functions and exclusive processes, NLO calculations for dijet production in DIS and hadronic collisions~\cite{Taels:2022tza,Caucal:2021ent,Iancu:2020mos,Boussarie:2016ogo}, and inclusive particle production in proton-nucleus collisions~\cite{Ducloue:2017dit,Ducloue:2016shw,Stasto:2013cha,Altinoluk:2014eka,Watanabe:2015tja,Iancu:2016vyg,Chirilli:2012jd,Liu:2020mpy,Liu:2022ijp} are becoming available.

The present paper completes the calculation of exclusive heavy vector meson production at next-to-leading order that we initialized in previous publications. First, the relativistic corrections suppressed by the squared quark velocity $\sim v^2 \as^0$ were determined in Ref.~\cite{Lappi:2020ufv}. Later, in Ref.~\cite{Mantysaari:2021ryb} we calculated the next-to-leading order corrections $\sim v^0 \as$ to longitudinally polarized heavy vector meson production. This paper presents the calculation of transversely polarized heavy vector meson in virtual photon-target scattering at the order $\as v^0$, and demonstrates how the NLO corrections and the relativistic corrections can both be included consistently. This development enables us to present the first calculation of exclusive \jpsi production at the order $\mathcal{O}(\as v^0,\as^0 v^2)$, and comparisons with the HERA and LHC data are presented in this paper.

This manuscript is structured as follows. The exclusive vector meson production process in the dipole picture is first presented in Sec.~\ref{sec:exclusive}. The next-to-leading order QCD corrections for the transverse heavy vector meson production are calculated in Sec.~\ref{sec:nlo_vm}. The implementation of relativistic (velocity) corrections is discussed in Sec.~\ref{sec:relativistic} before presenting numerical results in Sec.~\ref{sec:numerical} and conclusions in Sec.~\ref{sec:conclusions}.

\section{High-energy scattering in the dipole picture}
\label{sec:exclusive}

\subsection{Exclusive vector meson production}

The high-energy limit allows us to describe exclusive scattering in a factorized form where different parts of the process can be written independently. We work in a frame where the photon plus momentum $q^+$ is very large and it has no transverse momentum. The splitting of the virtual photon and the vector meson formation are described by the (boost invariant) light-front wave functions of the photon ($\Psi_{\gamma^*}$) and meson ($\Psi_V$).
At leading order the only contribution comes from the photon splitting into a quark-antiquark dipole. Additional Fock states have to be introduced at higher orders in $\as$, and at next-to-leading order one has to include a contribution from the photon splitting into a $q\bar q g$ state. The corresponding NLO scattering amplitude for vector meson production at $t \approx  -\Deltat^2=0$ can be written as
\begin{multline}
    \label{eq:im_amplitude}
    -i \acal = 2 \int \dd[2]{\xt_0} \dd[2]{\xt_1} \int \frac{\dd[]{z_0}\dd[]{z_1}}{(4\pi)^2} 4\pi\delta(z_0+z_1-1) \Psi_{\gamma^*}^{q \bar q} \Psi^{q \bar q*}_V N_{01} \\
    + 2\int \dd[2]{\xt_0} \dd[2]{\xt_1} \dd[2]{\xt_2} \int \frac{\dd[]{z_0}\dd[]{z_1}\dd[]{z_2}}{(4\pi)^3} 4\pi\delta(z_0+z_1+z_2-1) \Psi_{\gamma^*}^{q \bar q g } \Psi^{q \bar q g*}_V  N_{012}.
\end{multline}
Here $\xt_i$ are the transverse coordinates of the quark ($i=0$), the antiquark ($i=1$) and the gluon ($i=2$), and $z_i$ are the corresponding fractions of the photon plus momentum. The different helicity and color components of the wave functions are summed over implicitly. The coherent vector meson production cross section then reads~\cite{Good:1960ba}
\begin{equation}
\label{eq:vm_xs}
   \left.\frac{\der \sigma}{\der t}\right|_{t=0} = \frac{1}{16\pi} \left| \langle \acal  \rangle \right|^2.
\end{equation}

The action of the Wilson lines $V(\xt_i)$ on the quark-antiquark dipole is given by the dipole amplitude $N_{01}$:
\begin{equation}
\label{eq:S-matrix_01}
    1-N_{01} = \Re \frac{1}{\nc} \left\langle \Tr V(\xt_0) V^\dagger(\xt_1) \right\rangle.
\end{equation}
The dipole-target scattering amplitude $N_{01}$ depends  on the transverse separation $\xt_{01}=\xt_0-\xt_1$, impact parameter $\bt=(\xt_0+\xt_1)/2$ and projectile evolution rapidity $Y$. The evolution rapidity depends on the photon-nucleon system center-of-mass energy $W$ as discussed in Sec.~\ref{sec:evolution}.  The notation $\langle \cdots \rangle$ corresponds to the average of the target color charge configurations, which is done at the amplitude level in Eq.~\eqref{eq:vm_xs} when calculating coherent (i.e. no target dissociation) vector meson production~\cite{Good:1960ba} (see also e.g. Refs.~\cite{Mantysaari:2020axf,Mantysaari:2016ykx,Caldwell:2010zza} related to incoherent diffraction and discussion about the averaging procedure).  Similarly, interaction of the $q \bar q g$ system with the target is given in terms of the dipole amplitude $N_{012}$ which in the mean field limit can be written as~\cite{Hanninen:2017ddy}
\begin{equation}
	\label{eq:S-matrix_012}
	1-N_{012} = \frac{N_c}{2 C_F} \left( S_{02} S_{12} - \frac{1}{N_c^2} S_{01}\right),
\end{equation}
where $S_{ij}=1-N_{ij}$.

As the impact parameter is conjugate to the momentum transfer in the process, the impact parameter dependence of the dipole amplitudes can be connected to the $t$-dependence of the production amplitude. However, the impact parameter dependence of the dipole amplitude requires additional modeling and an effective description of confinement effects (see e.g.~\cite{Berger:2012wx,Berger:2011ew,Berger:2010sh,Mantysaari:2018zdd}), and for simplicity we choose to study only the case $t=0$ given by Eq.~\eqref{eq:im_amplitude} where only the impact parameter integrated dipole amplitude constrained by structure function measurements~\cite{Beuf:2020dxl} contributes.

In general, the production amplitude depends on both the polarization of the photon $\lambda_\gamma$ and the vector meson $\lambda_V$. The polarization mixing $\lambda_\gamma \neq \lambda_V$ is heavily suppressed and consequently it is sufficient to consider only the case $\lambda_\gamma = \lambda_V$ \cite{Mantysaari:2020lhf}.
Vector meson production can then be divided into longitudinal and transverse production, of which longitudinal channel has already been calculated at NLO by us in Ref.~\cite{Mantysaari:2021ryb}. In this paper we complete the NLO production calculation by computing the transverse production case, allowing us to consider total vector meson production.

\subsection{High-energy evolution}
\label{sec:evolution}

The center-of-mass energy or, equivalently, Bjorken-$x$ dependence of the Wilson lines can be obtained by solving the perturbative JIMWLK~\cite{Jalilian-Marian:1996mkd,Jalilian-Marian:1997qno, Jalilian-Marian:1997jhx,Iancu:2001md, Iancu:2001ad, Ferreiro:2001qy,Iancu:2000hn} evolution equation. In the large-$\nc$ limit one can derive from it the BK equation describing the energy (evolution rapidity $Y$) dependence of the dipole amplitude $N_{01}$:
\begin{equation}
    \label{eq:BK}
    \frac{\partial S_{01}}{\partial Y} = \int \dd[2] \xt_2 K_\text{BK}(\xt_0,\xt_1,\xt_2) [S_{02} S_{12}-S_{01}].
\end{equation}
The kernel $K_\text{BK}$ describes the probability to emit a gluon at the transverse position $\xt_2$ from the quark-antiquark dipole at the coordinates $\xt_0$ and $\xt_1$. Including the running-coupling corrections following~\cite{Balitsky:2006wa}, the kernel reads
\begin{equation}
    \label{eq:bk-rc-balitsky}
  K_\text{BK}(\xt_0,\xt_1, \xt_2) = \frac{\nc \as(\xt_{01}^2)}{2\pi^2} \left[
        \frac{\xt_{01}^2}{\xt_{21}^2 \xt_{20}^2}  + \frac{1}{\xt_{20}^2} \left( \frac{\as(\xt_{20}^2)}{\as(\xt_{21}^2)} -1 \right) 
        + \frac{1}{\xt_{21}^2} \left( \frac{\as(\xt_{21}^2)}{\as(\xt_{20}^2)} -1 \right)
  \right],
\end{equation}
where $\xt_{ij}=\xt_i-\xt_j$.

The BK equation at next-to-leading order, and a numerical solution to it, are available~\cite{Balitsky:2008zza,Lappi:2015fma,Lappi:2016fmu,Lappi:2020srm} (as well as the NLO JIMWLK equation~\cite{Balitsky:2013fea,Kovner:2013ona}). In principle it would be consistent to use the full NLO evolution equation when calculating vector meson production at this order in $\as$.
However, the NLO BK equation is numerically demanding due to an extra transverse integral, which is also the reason why there is no initial condition to it fitted to experimental data. In this work we follow Ref.~\cite{Beuf:2020dxl} and use the leading-order BK evolution equation combined with different implementations of a resummation of the most important higher-order contributions. These resummations are known to approximate the full NLO BK equation well as shown in Ref.~\cite{Lappi:2016fmu,Hanninen:2021byo}. The initial conditions for these evolutions are determined in  Ref.~\cite{Beuf:2020dxl} by performing a fit to HERA structure function data~\cite{Aaron:2009aa,Abramowicz:2015mha}. In our numerical analysis we use the fit results from publicly available codes~\cite{heikki_mantysaari_2020_4229269}. The running strong coupling constant in coordinate space is evaluated using the same parametrization as in the corresponding dipole amplitude fits in Ref.~\cite{Beuf:2020dxl}. The explicit expression for the running coupling is
\begin{equation}
    \label{eq:running_coupling}
    \as(\xt_{ij}^2)=\frac{4\pi}{\beta_0 \ln \left[ \left(\frac{\mu_0^2}{ \Lambda_\text{QCD}^2}\right)^{1/c}+\left(\frac{4C^2}{\xt_{ij}^2 \Lambda_\text{QCD}^2}\right)^{1/c} \right]^c}
\end{equation}
with $\Lambda_\text{QCD}= 0.241 \gev$, $c = 0.2$, $\mu_0/\Lambda_\text{QCD} = 2.5$, $\beta_0 = (11N_c- 2N_F)/3$ and $N_F=3$, and $C^2$ is a fit parameter determined when the initial condition for the BK evolution is fitted to the HERA data. 

The three different schemes to include resummation of higher-order corrections into the BK equation equation used in this work are, following the terminology of Ref.~\cite{Beuf:2020dxl}, \emph{KCBK}~\cite{Beuf:2014uia}, \emph{ResumBK}~\cite{Iancu:2015vea,Iancu:2015joa} and \emph{TBK}~\cite{Ducloue:2019ezk}. 
The evolution rapidity in the KCBK and ResumBK equations is related the fraction of the projectile (photon) plus momentum carrried by the gluon: 
\begin{equation} 
Y = \ln \frac{k^+}{P^+},
\end{equation} 
where $k^+=z_2 q^+$ is the gluon plus momentum and $P$ is the target momentum. The evolution rapidity in the TBK equation is related to the target longitudinal momentum fraction as we will discuss shortly.

The KCBK (``kinematically constrained BK equation'') is derived in Ref.~\cite{Beuf:2014uia} by requiring the necessary time ordering between the subsequent gluon emissions. This procedure effectively resums corrections that are enhanced by double transverse logarithm $\sim \as \ln \frac{\xt_{02}}{\xt_{01}} \ln \frac{\xt_{12}}{\xt_{01}}$. The same logarithms are included in the ResumBK (``resummed BK'') equation, with the difference that in Ref.~\cite{Iancu:2015vea} a form of the evolution equation which is local in rapidity $Y$ is derived.  Additionally, the ResumBK evolution equation further includes a resummation of single transverse logarithms~\cite{Iancu:2015joa} $\sim \as \ln \frac{1}{\xt_{ij}^2 \qs^2}$ to all orders. For explicit expressions for these evolution equations, see Ref.~\cite{Beuf:2020dxl}.

The third evolution equation used in this work is the TBK equation (``BK equation in target rapidity''), where the evolution rapidity $\eta$ is expressed in terms of the fraction of the target longitudinal (minus) momentum transferred in the process $\xpom$ (see detailed discussion in Ref.~\cite{Ducloue:2019ezk}):
\begin{equation}
\label{eq:xpom}
    \xpom \approx \frac{M_V^2 + Q^2}{W^2+Q^2},
\end{equation}
where $M_V$ is the meson mass.
Consequently the TBK evolution can be thought of as evolution in $\ln 1/\xpom$, whereas the KCBK and ResumBK evolutions written in terms of the projectile rapidity $Y$ are evolutions in $\ln W^2$~\cite{Ducloue:2019ezk,Beuf:2020dxl}.

When using a solution to the TBK evolution, written in terms of the target rapidity $\eta$, in the NLO impact factors calculated in this work that are written in terms of projectile rapidity $Y$ we use the same shift as in Ref.~\cite{Beuf:2020dxl}:
\begin{equation}
    \eta = Y - \ln \frac{1}{\min\{1, \xt_{01}^2 Q_0^2 \} },
\end{equation}
where the target transverse momentum scale is set to $Q_0^2=1\gev^2$.

Initial conditions for all these three evolution equations are obtained in Ref.~\cite{Beuf:2020dxl} by parametrizing the initial condition and fitting the free parameters to the HERA reduced cross section data. In this work we use the fit results obtained using the ``Balitsky $+$ smallest dipole'' running coupling scheme. We note that in Ref.~\cite{Beuf:2020dxl} only the light quark contribution is included in the NLO structure function calculations. On the other hand, in this work we consider heavy vector meson production, and as such it is not fully consistent to use the fit results from Ref.~\cite{Beuf:2020dxl}. However, the main purpose of this work is to derive the cross section at NLO accuracy, and detailed phenomenological comparisons to experimental data should be done later when the initial condition for the BK evolution is determined including the effect of quark masses.

\section{Vector meson production at next-to-leading order}
\label{sec:nlo_vm}

Next-to-leading order corrections to exclusive vector meson production consist of corrections from perturbative gluons. These can be included by calculating the virtual photon and meson wave functions at proper order in $\as$ such that we have all the corrections at the order $\as$ at the amplitude level. This means that we have to include the $\mathcal{O}(\as)$ loop corrections to the light-cone wave functions $\Psi^{q \bar q}_\gamma$ and $\Psi^{q \bar q}_V$, and also take into account the contribution from the $q \bar q g$ state with the wave functions $\Psi^{q \bar q g}_\gamma$ and $\Psi^{q \bar q g}_V$. The NLO wave function for the transverse photon with massive quarks has been calculated in Refs.~\cite{Beuf:2021srj,Beuf:2022ndu}, and the NLO heavy vector meson wave function in the nonrelativistic limit is evaluated in Ref.~\cite{Escobedo:2019bxn}. These results are applied in this work.

For completeness, we present here the next-to-leading order wave functions that enter our calculations. Our notation follows mostly Refs.~\cite{Beuf:2021srj,Beuf:2022ndu} with the exception that the integration measure is chosen to be $\prod_i \frac{\dd[2]{\xt_{i}}\dd{z_i}}{4\pi}$ where $i$ goes over the partons of the Fock state corresponding to the wave function. This introduces additional normalization factors $\frac{1}{2q^+}\prod_i \frac{1}{\sqrt{z_i}}$ compared to the photon wave functions presented in~\cite{Beuf:2021srj,Beuf:2022ndu}. 
Also, we choose to use the conventional dimensional regularization (CDR) scheme for our calculations, which corresponds to the case $D_s =D$ in Refs.~\cite{Beuf:2021srj,Beuf:2022ndu}.

The wave functions contain divergences that need to be regularized. Ultraviolet (UV) divergences are regularized using dimensional regularization in $D-2$ dimensions for the transverse coordinates. Infrared (IR) divergences originating from gluons with zero plus momenta are removed by introducing a cut-off $\alpha$ for the gluon plus momenta, $k_2^+ > \alpha q^+$ where $\alpha >0$ and $q^+$ is the plus momentum of the photon. The divergences will cancel in the calculation, and at the end we will take the limit $D \rightarrow 4$ and $\alpha \rightarrow 0$.

\subsection{Virtual photon wave function at next-to-leading order}
\label{sec:photon_wf}

 With these conventions, the LO transverse photon wave function for the $q \bar q$ state  (at zero photon transverse momentum, $\mathbf{q}=0$) is~\cite{Beuf:2022ndu}
\begin{equation}
    \begin{split}
	\label{eq:qq_LO_photon}
		\Psi_\lo^{\gamma^* \rightarrow q \bar q} =& -\frac{1}{2 q^+ \sqrt{z_0 (1-z_0)}} \frac{e e_f }{2 \pi} \left( \frac{\kappa_z}{2 \pi |\xt_{01}|} \right)^{(D-4)/2} \epsilon^j_{\lambda_\gamma} \delta_{\alpha_0 \alpha_1} \\
		\times  \Bigg\{&\bar u(0) \left[ \left( 2z_0-1  \right) \delta^{ij} \gamma^+ + \frac{1}{2}  \gamma^+ \left[ \gamma^i, \gamma^j \right]   \right]v(1)   i \kappa_z  \frac{\xt_{01}^i }{|\xt_{01}|} K_{(D-4)/2+1} \left( |\xt_{01}| \kappa_z\right) \\
		&- m_q \bar u(0) \gamma^+ \gamma^j v(1) K_{(D-4)/2} \left( |\xt_{01}| \kappa_z\right) \Bigg\}
    \end{split}
\end{equation}
where $u(0)$ and $v(1)$ are spinors corresponding to the quark and antiquark, $m_q$ is the heavy quark mass, $\kappa_v = \sqrt{ v (1-v) Q^2+m_q^2}$ where $Q^2$ is the photon virtuality, and $\alpha_0, \alpha_1$ are the color indices of the quark and antiquark. The fraction of the photon plus momentum carried by the quark is $z_0$. Repeated indices in the Latin alphabet are summed over in $D-2$ transverse dimensions. The functions $K_\nu$ are modified Bessel functions of the second kind. The NLO correction to the $q \bar q$ wave function is ~\cite{Beuf:2021srj,Beuf:2022ndu}
\begin{equation}
	\label{eq:qq_NLO_photon}
	\begin{split}
		\Psi_\nlo^{\gamma^* \rightarrow q \bar q} =& -\frac{1}{2 q^+ \sqrt{z_0 (1-z_0)}} \frac{e e_f}{2\pi} \left( \frac{\as C_F}{2 \pi}\right) \epsilon^j_{\lambda_\gamma} \delta_{\alpha_0 \alpha_1} \\
		\times \Bigg\{& \bar u(0)\left[\left( 2z_0-1  \right) \delta^{ij}  \gamma^+  + \frac{1}{2} \gamma^+ \left[\gamma^i, \gamma^j \right] \right] v(1) \fcal[\Pt^i \vcal^T] \\
		& + \bar u(0)  \gamma^+ v(1) \fcal[\Pt^j \ncal^T]+m_q \bar u(0) \gamma^+ \gamma^i v(1) \fcal \left[ \left( \frac{\Pt^i \Pt^j}{\Pt^2} - \frac{1}{2}\delta^{ij} \right) \scal^T \right] \\
		&- m_q \bar u(0) \gamma^+ \gamma^j v(1) \fcal \left[ \mcal^T+\vcal^T - \frac{1}{2} \scal^T \right] \Bigg\},
	\end{split}
\end{equation}
where the wave function is written in terms of different form factors. It will turn out that we do not need to know the explicit expressions for all of these form factors, which follows from the fact that at leading order in $\as$ and $v$ the vector meson spin structure is very simple, picking up only parts with an odd number of transverse gamma matrices. 
In addition to this, the traceless part $ \frac{\Pt^i \Pt^j}{\Pt^2} -\frac{1}{2} \delta^{ij} $ also vanishes as after taking the gamma matrix traces one gets $\epsilon_{\lambda_\gamma}^j \epsilon_{\lambda_V}^{i*} \left(  \frac{\Pt^i \Pt^j}{\Pt^2} -\frac{1}{2} \delta^{ij}\right) =0$ which is valid for $\lambda_\gamma = \lambda_V$ in 2 transverse dimensions. In the polarization mixing case, $\lambda_\gamma \neq \lambda_V$, there would be a non-zero contribution from this term.
Thus, only the last term in Eq.~\eqref{eq:qq_NLO_photon} contributes to vector meson production in the nonrelativistic limit, and the required combination of the form factors reads
\begin{equation}
	\label{eq:FT_MVS}
	\begin{split}
		&\fcal \left[\mcal^T + \vcal^T - \frac{1}{2}\scal^T \right] =
		 \left( \frac{\kappa_z}{2\pi |\xt_{01}|} \right)^{(D-4)/2} K_{(D-4)/2} \left(  |\xt_{01}| \kappa_z \right)\Bigg\{\frac{1}{2}\\
		&+\left[ \frac{3}{2}+ \ln(\frac{\alpha}{z_0})+ \ln(\frac{\alpha}{1-z_0}) \right]\left[ \frac{2 \left( 4 \pi\right)^{(4-D)/2}}{4-D} \Gamma\left(1+\frac{4-D}{2}\right)+\ln(\frac{\xt_{01}^2 \mu^2}{4})+2 \gamma_E  \right] \Bigg\} \\
		&+  K_{0} \left(  |\xt_{01}| \kappa_z \right) \left\{\Omega^T_\vcal(\gamma;z_0) + L(\gamma;z_0)- \frac{\pi^2}{3} + \ln^2 \left(\frac{z_0}{1-z_0 }\right)+3 \right\} \\ 
		&+ \widetilde I^T_{\vcal \mcal \scal}(|\xt_{01}|, z_0).
	\end{split}
\end{equation}
Here $\mu$ is the mass scale coming from dimensional regularization. The functions $\Omega^T_\vcal$ and $L$ are defined as
\begin{equation}
	\label{eq:Omega_V}
	\begin{split}
	\Omega^T_\vcal(\gamma;z) =& \left(1 + \frac{1}{2z}\right) \left[ \ln(1-z)+\gamma \ln\left(  \frac{1+\gamma}{1+\gamma-2z} \right) \right] \\
	&-\frac{1}{2z} \left[ \left(z+\frac{1}{2}\right) (1-\gamma)+\frac{m_q^2}{ Q^2} \right] \ln( \frac{ \kappa_z^2 }{m_q^2} ) \\
	&+\left( z \leftrightarrow 1-z \right),
	\end{split}
\end{equation}
and 
\begin{equation}
	\label{eq:L}
	L(\gamma;z) = \sum_{\sigma = \pm 1} \left[ \text{Li}_2 \left(\frac{1}{1-\frac{1}{2z}(1+\sigma\gamma)}\right)
	+\text{Li}_2 \left(\frac{1}{1-\frac{1}{2(1-z)}(1+\sigma\gamma)}\right) \right]
\end{equation}
where $\text{Li}_2$ is the dilogarithm function and
\begin{equation}
	\label{eq:gamma}
	\gamma = \sqrt{1+\frac{4m_q^2}{Q^2}}.
\end{equation}
The function $\widetilde I^T_{\vcal \mcal \scal}(r, z)$ can be written in the form 
\begin{equation}
    \label{eq:I_VMS}
    \begin{split}
        	\widetilde I^T_{\vcal \mcal \scal}&(r,z) =\\
        	\int_0^1 \dd[]{\xi}& \Bigg\{
	        \frac{1}{\xi}\left[\frac{2\ln \xi}{1-\xi} -\frac{1+\xi}{2}\right] \left[ K_0\left( r\sqrt{\kappa_z^2 +\frac{\xi(1-z)}{1-\xi}m_q^2} \right) - K_0\left( r \kappa_z \right) \right] \\
	        &+\left[-\frac{3(1-z)}{2(1-\xi)}+\frac{1-z}{2} \right]K_0\left( r\sqrt{\kappa_z^2 +\frac{\xi(1-z)}{1-\xi}m_q^2} \right)\Bigg\} \\
	        +\int_0^z \dd{\chi}& \int_0^\infty \! \! \dd{u} \Bigg\{ 
	        \frac{1}{1-\chi}\frac{1}{(u+1)^2} \left[ -z-\frac{u}{1+u}\frac{z+u\chi}{z} \left( \chi-(1-z)\right) \right] K_0\left( r \sqrt{ \kappa_z^2 + u \frac{1-z}{1-\chi} \kappa_\chi^2} \right)\\
	        &+\frac{1}{(u+1)^3} \left[ \frac{\kappa_z^2}{\kappa_\chi^2} \left( 1+ u\frac{\chi(1-\chi)}{z(1-z)} \right) - \frac{m_q^2}{\kappa_\chi^2} \frac{\chi}{1-\chi} \left( 2\frac{(1+u)^2}{u}+ \frac{u}{z(1-z)}  \left( z-\chi \right)^2 \right) \right]\\
	        &\quad\quad \times \left[  K_0\left( r \sqrt{\kappa_z^2 + u \frac{1-z}{1-\chi} \kappa_\chi^2}\right)-K_0 \left( r \kappa_z\right) \right]\Bigg\} \\
	        &+ \left(z \leftrightarrow 1-z\right),
    \end{split}
\end{equation}
where the substitution $ ( z \leftrightarrow 1-z) $ corresponds to the whole expression.

In addition to the $q \bar q$ wave function of the photon, we also need the light-front wave function for the $q \bar q g$ state (again for the photon with zero transverse momentum). This can written as~\cite{Beuf:2022ndu}
\begin{equation}
	\label{eq:qqg_photon}
		\psi^{\gamma^* \rightarrow q \bar qg} = \frac{1}{2q^+ \sqrt{z_0z_1z_2}} t^a_{\alpha_0 \alpha_1} e e_f g \epsilon_{\lambda_\gamma}^l \epsilon_\sigma^{*j} \left( \Sigma^{lj} + \Sigma^{l j}_m \right),
\end{equation}
where $\sigma$ is the gluon helicity and
\begin{equation}
    \begin{split}
	\label{eq:sigma_lj}
		\Sigma^{lj} =& -\frac{1}{(z_0 + z_2) } \bar u(0) \gamma^+  \left[ (2z_0  + z_2) \delta^{ij}-\frac{z_2}{2} \left[ \gamma^i, \gamma^j \right] \right] \\
		&\hspace{1cm}\times\left[ (2z_1 -1) \delta^{kl}-\frac{1}{2} \left[ \gamma^k, \gamma^l \right] \right] v(1)\ical_{(j)}^{ik} \\
		&-\frac{1}{(z_1 + z_2) } \bar u(0) \gamma^+ \left[ (2z_0 -1) \delta^{kl}+\frac{1}{2} \left[ \gamma^k, \gamma^l \right] \right]\\
		&\hspace{1cm}\times\left[ (2z_1  + z_2) \delta^{ij}+\frac{z_2}{2} \left[ \gamma^i, \gamma^j \right] \right] v(1)\ical_{(k)}^{ik} \\
		&+ \frac{z_2 z_0}{(z_0+z_2)^2} \bar u(0) \gamma^+ \gamma^j \gamma^l v(1) \jcal_{(l)} - \frac{z_2 z_1}{(z_1 + z_2)^2} \bar u(0) \gamma^+ \gamma^l \gamma^j v(1) \jcal_{(m)}
    \end{split}
\end{equation}
and 
\begin{equation}
	\label{eq:sigma_ljm}
	\begin{split}
		\Sigma^{lj}_m =& -m_q\frac{1}{(z_0 + z_2)} \bar u(0)\gamma^+  \left[ (2z_0  + z_2) \delta^{ij}-\frac{z_2}{2} \left[ \gamma^i, \gamma^j \right] \right]  \gamma^l v(1) \ical_{(j)}^i\\
		& +m_q\frac{z_2^2}{(z_0 + z_2)^2 } \bar u(0)\gamma^+ \gamma^j\left[ (2z_1 -1) \delta^{kl}-\frac{1}{2} \left[ \gamma^k, \gamma^l \right] \right]v(1) \hat \ical_{(j)}^k\\
		& +m_q^2\frac{z_2^2}{(z_0 + z_2)^2} \bar u(0)\gamma^+ \gamma^j \gamma^l v(1) \ical_{(j)}\\
		& +m_q\frac{1}{(z_1 + z_2)} \bar u(0)\gamma^+ \gamma^l  \left[ (2z_1  + z_2) \delta^{ij}+\frac{z_2}{2} \left[ \gamma^i, \gamma^j \right] \right] v(1) \ical_{(k)}^i\\
		& +m_q\frac{z_2^2}{(z_1 + z_2)^2 } \bar u(0)\gamma^+ \left[ (2z_0 -1) \delta^{kl}+\frac{1}{2} \left[ \gamma^k, \gamma^l \right] \right] \gamma^j v(1) \hat \ical_{(k)}^k\\
		& -m_q^2\frac{z_2^2}{(z_1 + z_2)^2} \bar u(0)\gamma^+ \gamma^l \gamma^j v(1) \ical_{(k)}.
	\end{split}
\end{equation}
The special functions use the following labeling for the subindices
\begin{align}
    \ical_{(j)}&=\ical(\xt_{0+2;1},\xt_{20}, \overline Q^2_{(j)}, \omega_{(j)}, \lambda_{(j)}) \quad &\ical_{(k)}=\ical(\xt_{0;1+2},\xt_{21}, \overline Q^2_{(k)}, \omega_{(k)}, \lambda_{(k)}) \\
    \hat \ical_{(j)}&=\hat \ical(\xt_{0+2;1},\xt_{20}, \overline Q^2_{(j)}, \omega_{(j)}, \lambda_{(j)}) \quad &\hat \ical_{(k)}=\hat \ical(\xt_{0;1+2},\xt_{21}, \overline Q^2_{(k)}, \omega_{(k)}, \lambda_{(k)}) \\
    \jcal_{(l)}&=\jcal(\xt_{0+2;1},\xt_{20}, \overline Q^2_{(j)}, \omega_{(j)}, \lambda_{(j)}) \quad &\jcal_{(m)}=\jcal(\xt_{0;1+2},\xt_{21}, \overline Q^2_{(k)}, \omega_{(k)}, \lambda_{(k)}) 
\end{align}
(and analogously for the special functions with transverse indices), where
\begin{equation}
\begin{aligned}
	\omega_{(j)} &= \frac{ z_0 z_2}{z_1(z_0 + z_2)^2}, &
	\omega_{(k)} &= \frac{ z_1 z_2}{z_0(z_1 + z_2)^2}, \\
	\overline Q^2_{(j)} &= z_1 (1-z_1) Q^2, &
	\overline Q^2_{(k)} &= z_0 (1-z_0) Q^2, \\
	\lambda_{(j)} &= \frac{z_1 z_2}{z_0}, &
	\lambda_{(k)} &= \frac{z_0 z_2}{z_1}, \\
\end{aligned}
\end{equation}
\begin{equation}
	\label{eq:x_nmp}
	\xt_{n+m;p} = - \xt_{p;n+m} = \frac{z_n \xt_n+z_m \xt_m}{z_n+z_m} - \xt_p.
\end{equation}
The special functions $\ical$ are defined as:
\begin{multline}
	\label{eq:I_ij}
	\ical^{ij}(\bt,\rt,\overline Q^2, \omega, \lambda) =\\
	 -\frac{\mu^{(4-D)/2}}{4 (4\pi)^{D-2}} \bt^i \rt^j \int_0^\infty \dd[]{u} u^{-D/2} e^{-u(\overline Q^2+m_q^2)} e^{-\bt^2/(4u)} \int_0^{u/\omega} \dd[]{t} t^{-D/2} e^{-t \omega \lambda m_q^2} e^{-\rt^2/(4t)},
\end{multline}
\begin{multline}
	\label{eq:I_i}
	\ical^i(\bt,\rt,\overline Q^2, \omega, \lambda) = \\
	\frac{i \mu^{(4-D)/2}}{2 (4\pi)^{D-2}} \rt^i \int_0^\infty \dd[]{u} u^{1-D/2} e^{-u(\overline Q^2+m_q^2)} e^{-\bt^2/(4u)} \int_0^{u/\omega} \dd[]{t} t^{-D/2} e^{-t \omega \lambda m_q^2} e^{-\rt^2/(4t)},
\end{multline}
\begin{multline}
	\label{eq:I_hat_i}
	\hat \ical^i(\bt,\rt,\overline Q^2, \omega, \lambda) =\\
	\frac{i \mu^{(4-D)/2}}{2 (4\pi)^{D-2}} \bt^i \int_0^\infty \dd[]{u} u^{-D/2} e^{-u(\overline Q^2+m_q^2)} e^{-\bt^2/(4u)} \int_0^{u/\omega} \dd[]{t} t^{1-D/2} e^{-t \omega \lambda m_q^2} e^{-\rt^2/(4t)},
\end{multline}
\begin{multline}
	\label{eq:I}
	\ical(\bt,\rt,\overline Q^2, \omega, \lambda) = \\
	\frac{ \mu^{(4-D)/2}}{ (4\pi)^{D-2}} \int_0^\infty \dd[]{u} u^{1-D/2} e^{-u(\overline Q^2+m_q^2)} e^{-\bt^2/(4u)} \int_0^{u/\omega} \dd[]{t} t^{1-D/2} e^{-t \omega \lambda m_q^2} e^{-\rt^2/(4t)},
\end{multline}
\begin{multline}
	\label{eq:J}
	\jcal(\bt,\rt,\overline Q^2, \omega, \lambda) =\\
	(2 \pi)^{2-D} \left( \frac{\mu}{\omega}\right)^{\frac{4-D}{2}} \left(\sqrt{\frac{\overline Q^2+ m_q^2 + \lambda  m_q^2}{\bt^2 + \omega \rt^2}}\right)^{D-3} K_{D-3}\left(\sqrt{\overline Q^2 + m_q^2 + \lambda m_q^2} \sqrt{ \bt^2+ \omega \rt^2 }  \right).
\end{multline}

\subsection{Meson wave function at next-to-leading order}

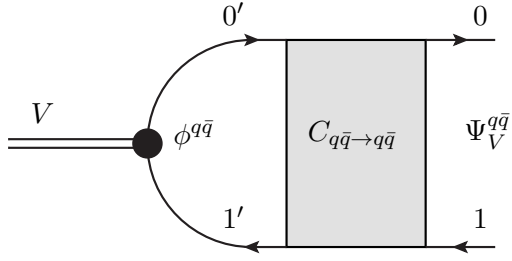
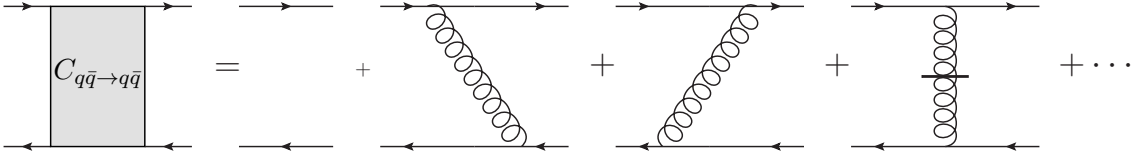
\begin{figure}
	\centering
    \begin{subfigure}{\textwidth}
        \centering
        \includegraphics[height=0.14\textheight]{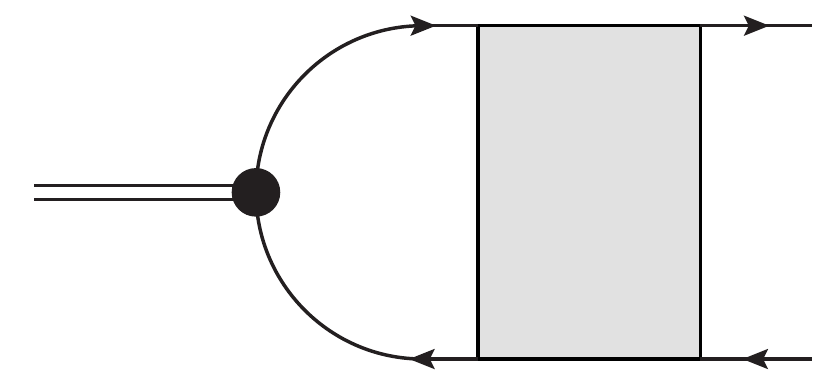}
        \begin{tikzpicture}[overlay]
         \node[anchor=south east] at (-3.5cm,3.0cm) {$0'$};
         \node[anchor=south east] at (-3.5cm,0.3cm) {$1'$};
         \node[anchor=south east] at (-0.3cm,3.0cm) {$0$};
         \node[anchor=south east] at (-0.3cm,0.3cm) {$1$};
         \node[anchor=south east] at (-1.5cm,1.35cm) {$C_{q \bar q \rightarrow q \bar q}$};
         \node[anchor=south east] at (0cm,1.35cm) {$\Psi_V^{q \bar q}$};
         \node[anchor=south east] at (-3.9cm,1.35cm) {$\phi^{q \bar q}$};
         \node[anchor=south east] at (-6.0cm,1.7cm) {$V$};         \end{tikzpicture}
        \caption{ Contribution of the leading-order wave function $\phi^{q\bar q}$ to the total vector meson  wave function. }
        \label{fig:C_qq_qq}
    \end{subfigure}
    \begin{subfigure}{\textwidth}
        \centering
        \includegraphics[width=\textwidth]{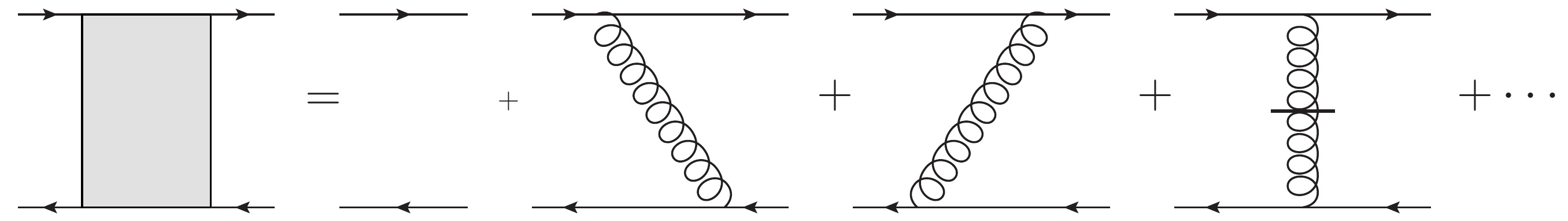}
        \begin{tikzpicture}[overlay]
         \node[anchor=south east] at (-5.43cm,1.3cm) {$C_{q \bar q \rightarrow q \bar q}$};      
         \end{tikzpicture}
        \caption{ The coefficient function $C_{q \bar q \leftarrow q \bar q}$ in terms of Feynman diagrams. The self-energy corrections are not shown. }
        \label{fig:C_qq_qq-kernel}
    \end{subfigure}
	  \caption{Perturbative corrections to the meson light-front wave function $\Psi^{q\bar q}_V$. 
	  }
        \label{fig:nonrelativistic_expansion}
\end{figure}

For the heavy vector meson wave function we can use the nonrelativistic expansion developed in Ref.~\cite{Escobedo:2019bxn}. In this expansion, corrections in $\as$ are included in factors multiplying the leading-order wave function, and corrections suppressed by the heavy quark velocity $v$ appear as derivatives of the leading-order wave function. For a general Fock state $n$ we write this in the following form:
\begin{equation}
	\label{eq:NR_expansion}
	\Psi^n_V = \sum_{m,k} C^k_{n \leftarrow m} \int_0^1 \frac{\dd{z'}}{4\pi}  \left(\frac{1}{m_q}\nabla\right)^k\phi^m(\rt = 0,z'),
\end{equation}
where $\phi^m$ is the leading-order wave function (LOWF) for the Fock state $m$ and the sum goes over all Fock states $m$. The case $m= q\bar q$, $n=q\bar q$ is shown schematically in Fig.~\ref{fig:nonrelativistic_expansion}, where the primed indices $0'$, $1'$ correspond to the nonrelativistic quark and antiquark in the LOWF and the non-primed indices $0$, $1$ correspond to the quark and antiquark in the wave function $\Psi^{q \bar q}$.
It should be noted that the coefficient functions $C^k_{n \leftarrow m} $ depend on  colors and helicities of the particles in Fock states $n$ and $m$, and the sum over these is left implicit. The derivative $\nabla$ is defined in the mixed space as $ \nabla = (\partial_{r_1},\partial_{r_2},(z'-1/2)2m_q i) $, where $r_i$ are the components of the transverse separation $\rt$, and $k=(k_1,k_2,k_3)$ is to be understood as a multi-index: $\left(\frac{1}{m_q}\nabla\right)^k = \frac{1}{m_q^{|k|}} \nabla_1^{k_1} \nabla_2^{k_2} \nabla_3^{k_3}$ where $|k|=k_1+k_2+k_3$.

The strength of this expansion is that it is now straightforward to include corrections at a given order in $\as$ and $v$. Higher-order corrections in $\as$ can be calculated perturbatively from Feynman diagrams, and they are defined as parts of the coefficient functions $C^k_{n \leftarrow m}$. Relativistic corrections in $v$ can be read from the derivatives which amount to a suppression of $v^{|k|}$.

At next-to-leading order in the nonrelativistic limit, we need the following wave functions \cite{Escobedo:2019bxn}:
\begin{equation}
	\label{eq:PsiV_qq}
	\Psi^{q \bar q}_V = C^{(0,0,0)}_{q \bar q \leftarrow q \bar q} \int_0^1 \frac{\dd{z'}}{4\pi}  \phi^{q \bar q}_{h_{0}' h_{1}'}(\rt = 0,z'),
\end{equation}
\begin{equation}
	\label{eq:PsiV_qqg}
	\Psi^{q \bar q g}_V = C^{(0,0,0)}_{q \bar q g \leftarrow q \bar q} \int_0^1 \frac{\dd{z'}}{4\pi}\phi^{q \bar q}_{h_{0}' h_{1}'}(\rt = 0,z').
\end{equation}
Here $C^{(0,0,0)}_{q \bar q \leftarrow q \bar q}$ is calculated to the order $\ocal(g^2)$ and $C^{(0,0,0)}_{q \bar q g \leftarrow q \bar q}$ to the order $\ocal(g)$ in the strong coupling constant.
Only the LOWF for the dominating Fock state $q \bar q$ contributes at this order in the expansion, as soft gluons in the LOWF Fock state would bring additional suppression in velocity (see also discussion in Ref.~\cite{Escobedo:2019bxn}). Relativistic corrections at leading order in $\as$ are discussed in more detail in Sec.~\ref{sec:relativistic}.

As the functions $C^{(0,0,0)}_{n \leftarrow m} $ are fully perturbative (and calculated at NLO accuracy in Ref.~\cite{Escobedo:2019bxn}), the nonperturbative physics is contained in the constants $\int\frac{\dd{z'}}{4\pi}\phi_{h_0' h_1'}^{q \bar q}(\rt = 0,z')$ where $h_0'$ and $h_1'$ are the helicities of the quark and antiquark in the LOWF.
The nonrelativistic limit requires that the helicity structure of LOWF is simply $\delta_{(h_0'+h_1')/2, \lambda_V}$ where $\lambda_V$ is the polarization of the vector meson.
To write the expressions in a more compact form we extract the helicity and color structure from the LOWF in the coefficients $\widetilde C^k_{n \leftarrow q \bar q}$:
\begin{equation}
	\label{eq:C_tilde}
	\widetilde C^{(0,0,0)}_{n \leftarrow q \bar q} = \frac{1}{\sqrt{N_c}} \delta_{\alpha_0 \alpha_1} \frac{-1}{m_q 2 \sqrt{2}} \bar u(0') \slashed{\epsilon}_{\lambda_V} v(1')C^{(0,0,0)}_{n \leftarrow q \bar q} 
\end{equation}
and write the spin-independent part of the LOWF as $\int \frac{\dd{z'}}{4\pi}  \phi^{q \bar q}(\rt = 0,z')$. The spinors $u(0'), v(1')$ correspond to the nonrelativistic quark and antiquark in the LOWF with $z_{0'} = z_{1'} = \frac{1}{2}$.
The perturbative coefficients required for the transverse NLO calculation in the nonrelativistic limit are then
\begin{equation}
	\label{eq:C0_qq}
	\begin{split}
		&\widetilde C^{(0,0,0)}_{q \bar q \leftarrow q \bar q} = \frac{1}{\sqrt{N_c}} \delta_{\alpha_0 \alpha_1}  \frac{1}{4\sqrt{k_0^+ k_1^+k_{0'}^+ k_{1'}^+} } \bar u(0)\gamma^+ u(0')\bar v(1')\gamma^+  v(1) \epsilon_{\lambda_V}^j
		\\
		\times & \Bigg\{ 4\pi \delta(z-1/2)(1+\delta Z)  \frac{1}{m_q 2\sqrt{2}} \bar u(0') \gamma^j v(1') 
		\\
		&+\frac{\alpha_s C_F}{2\pi}\frac{16\pi z_0(1-z_0)}{(z_0-1/2)^2} \frac{1}{m_q 2\sqrt{2}} \bar u(0') \gamma^j v(1')
		\left[\theta\left(z_0-\frac{1}{2}-\alpha\right)+ \theta\left(\frac{1}{2}-z_0-\alpha\right) \right] \\
		 &\quad  \times\left[K_0(\tau) -\frac{1}{2z_0(1-z_0)}\left( \theta \left(z_0-\frac{1}{2}\right)(1-z_0)+\theta\left(\frac{1}{2}-z_0\right)z_0\right) 
		\left( K_0(\tau)-\frac{\tau}{2}K_1(\tau)\right)\right] \\
		&-\frac{\alpha_s C_F}{2\pi} 8\pi i  K_0(\tau)
		\left( \theta \left(z_0-\frac{1}{2}-\alpha\right)(1-z_0)+\theta\left(\frac{1}{2}-z_0-\alpha\right)z_0\right) \\
		& \quad \times \frac{2m_q}{\sqrt{2} q^+} \xt_{01}^i \bar u(0') \left[\delta^{ij} \gamma^+ (2z_0-1)+\frac{1}{2}\gamma^+ \left[\gamma^i, \gamma^j \right] \right] v(1')
		\Bigg\}
	\end{split}
\end{equation}
and
\begin{equation}
	\label{eq:C0_qqg}
	\begin{split}
		\widetilde C_{q \bar q g \leftarrow q \bar q}^{(0,0,0)} 
		=&\frac{1}{\sqrt{N_c}} t^a_{\alpha_0 \alpha_1} \frac{g}{2\pi}  \frac{1}{(q^+)^2} \epsilon_\sigma^{j*} \bar u(0') \gamma^k \epsilon_{\lambda_V}^k v(1')  \bar v(1') \gamma^+ v(1) \\
		&\times
		\left( \frac{m_q z_2}{\pi |\xt_{20}| \mu}\right)^{(D-4)/2}\sqrt{\frac{2z_2}{1-2z_2}}\cdot 4\pi \delta(z_1-1/2)\\
		&\times \Bigg\{  -i(1-z_2)\frac{\xt_{20}^i}{ |\xt_{20}|} K_{D/2-1}\big(2m_q z_2 |\xt_{02}|\big) \bar u(0) \left[\delta^{ij}-\frac{z_2}{2(1-z_2)} [\gamma^i,\gamma^j]\right]\gamma^+ u(0')  \\
		&\quad \quad+z_2 K_{D/2-2}\big(2m_q z_2 |\xt_{20}|\big) \bar u(0) \gamma^j \gamma^+ u(0')\Bigg\} \\
		&+ \text{gluon emission from antiquark},
	\end{split}
\end{equation}
with
\begin{equation}
	\label{eq:tau}
	\tau = 2m_q |\xt_{01}| \left|z_0- \frac{1}{2} \right|.
\end{equation}
The wave function renormalization factor $\delta Z$ is calculated in the pole-mass scheme and is given by
\begin{multline}
	\label{eq:delta_Z}
	\delta Z = -\frac{\as C_F}{2\pi} \left[ \frac{1}{D-4}(4\ln(2\alpha)+3)+2\ln(2\alpha) \left(\ln(\frac{m_q^2}{4\pi\mu^2})+1+\ln(2\alpha)\right) \right. \\
	\left. +(4\ln(2\alpha)+3)\frac{\gamma_E}{2}+\frac{3}{2}\ln(\frac{m_q^2}{4\pi\mu^2})-2\right].
\end{multline}
The contribution from an antiquark emitting the gluon is left implicit in Eq.~\eqref{eq:C0_qqg}. Its contribution to the final expression is equal to the quark emitting the gluon, meaning that we can include its contribution by multiplying the result from the quark contribution by two.

\subsection{Calculation of the next-to-leading order production}
\label{sec:nlo_calculation}
The next-to-leading order amplitude \eqref{eq:im_amplitude} has contributions from two terms: the quark-antiquark dipole contribution $2 \int_{\xt_0 \xt_1} \int \frac{\dd[]{z_0}\dd[]{z_1}}{(4\pi)^2} 4\pi \delta(z_0+z_1-1)  \Psi_{\gamma^*}^{q \bar q}\Psi^{q \bar q*}_V N_{01}$ and real gluon emission $2 \int_{\xt_0 \xt_1 \xt_2} \int \frac{\dd[]{z_0}\dd[]{z_1} \dd[]{z_2}}{(4\pi)^3} 4\pi \delta(z_0+z_1+z_2-1) \Psi_{\gamma^*}^{q \bar q g} \Psi^{q \bar q g*}_V N_{012}$. The next-to-leading order corrections to the dipole term come from virtual gluon loops. Using the above expressions for the photon and meson wave functions it is possible to calculate these two contributions.

The evaluation of the dipole term yields
\begin{equation}
	\label{eq:dipole_part}
	\begin{split}
		&2 \int_{\xt_0 \xt_1} \int \frac{\dd[]{z_0}\dd[]{z_1}}{(4\pi)^2} 4\pi \delta(z_0+z_1-1)\Psi_{\gamma^*}^{q \bar q} \Psi^{q \bar q*}_V  N_{01}  \\
		&= \int_0^1 \frac{\dd{z'}}{4\pi}\phi^{q \bar q}(\rt = 0,z')\int \dd[D-2]{\xt_{01}} \dd[2]{\bt} N_{01}(\rt, \bt) \sqrt{\frac{N_c}{2}} \frac{2e e_f m_q}{\pi} \Bigg\{ K_0(\zeta)\\
		&+\frac{\alpha_s C_F}{2\pi} \Bigg\{  -\frac{2}{D-4} \left[ 4\ln(2\alpha) + 3 \right] K_{(D-4)/2} (\zeta)+ \widetilde I^T_{\vcal \mcal \scal}\left(\frac{1}{2}, \xt_{01}\right)+\kcal^T\\
		&\quad\quad+K_0(\zeta)\Bigg[ \Omega^T_\vcal\left(\gamma;\frac{1}{2}\right) + L\left(\gamma;\frac{1}{2}\right)- \frac{\pi^2}{3} +\frac{7}{2}\\
		&\quad\quad\quad\quad+\frac{1}{\alpha}
		+ 4\ln(2\alpha) \ln(\frac{2\pi^2 |\xt_{01}|^4 \mu^2 e^{\gamma_E}}{\zeta})+3\ln(\frac{4\pi^2|\xt_{01}|^3\mu^2}{m_q \zeta})  \Bigg]
			\Bigg\} \Bigg\}
	\end{split}
\end{equation}
where $\zeta = |\xt_{01}| \sqrt{\frac{1}{4}Q^2+ m_q^2}$, $\bt=\frac{1}{2}(\xt_0+\xt_1)$ and
\begin{multline}
	\label{eq:KT}
		\kcal^T=\int^{1/2}_0\dd[]{z}\Bigg\{ 4z|\xt_{01}|\kappa_z K_1\left(|\xt_{01}| \kappa_z \right) K_0(\tau) \left[1+(1-2z)^2\right]\\
		+\frac{1}{(z-1/2)^2}\Bigg\{8z(1-z)K_0\left(|\xt_{01}|\kappa_z \right) \left[K_0(\tau)-\frac{1}{2(1-z)}\left( K_0(\tau)-\frac{\tau}{2}K_1(\tau) \right)\right] \\
		-2K_0(\zeta)\left[ \frac{1}{2}+(z-1/2) \left( 1+2\gamma_E +2\ln(\frac{\tau}{2}) \right) \right] \Bigg\}\Bigg\}.
\end{multline}
For simplicity, we have chosen the LOWF to be real. The real emission contribution is
\begin{equation}
    \begin{split}
	\label{eq:real_emission_part}
		&2 \int_{\xt_0 \xt_1 \xt_2} \int \frac{\dd[]{z_0}\dd[]{z_1} \dd[]{z_2}}{(4\pi)^3} 4\pi \delta(z_0+z_1+z_2-1) \Psi_{\gamma^*}^{q \bar q g}\Psi^{q \bar q g*}_V  N_{012} \\
		=&\int_0^1 \frac{\dd{z'}}{4\pi}\phi^{q \bar q}(\rt = 0,z')  \int \dd[D-2]{\xt_{01}}\dd[D-2]{\xt_{20}} \dd[2]{\bt} \int_\alpha^{1/2}\dd[]{z_2} \\
		&\times N_{012} \sqrt{\frac{N_c}{2}} \frac{2e e_f m_q }{\pi} \frac{\alpha_s C_F}{2\pi} 4\pi\left(\frac{m_q z_2}{\pi |\xt_{20}| \mu}\right)^{\frac{D-4}{2}}  \\
		&\times\Bigg\{8 K_{(D-2)/2}(2m_q z_2 |\xt_{20}|) \frac{i \xt_{20}^i}{|\xt_{20}|}m_q
		    \Bigg[- \ical_{(j)}^i \left((D-2)z_2^2-2z_2+1 \right) - 
		    z_2^2\hat \ical_{(j)}^i  \left(2z_2-1 \right) \\
		&\hspace{5.5cm}+\ical_{(k)}^i \frac{1}{2z_2+1}  
		    +z_2^2\hat \ical_{(k)}^i \frac{1}{(2z_2+1)^2}\left( 4z_2^2-4z_2+1 \right)  \Bigg]\\
		&+8z_2 K_{0}\left( 2m_q z_2|\xt_{20}| \right) 
		 \Bigg[ \frac{1}{2}\ical_{(j)}^{ii} \left(-1+2z_2\right)
		    +\frac{1}{2(1+2z_2)}\ical_{(k)}^{ii} \left( 4z_2^2 +4z_2 +1\right)\\
		&\hspace{8cm}-2\left( (1-2z_2)z_2 \jcal_{(l)} +2m_q^2 z_2^2 \ical_{(j)} \right)
		    \Bigg] 
		    \Bigg\},
    \end{split}
\end{equation}
where we have set $D \rightarrow 4$ wherever possible and $\bt = z_0 \xt_0 + z_1 \xt_1 + z_2 \xt_2$.
Both the dipole term~\eqref{eq:dipole_part} and real gluon emission contributions~\eqref{eq:real_emission_part} have divergences in the $D\rightarrow 4$ (UV) and $\alpha \rightarrow 0$ (IR) limits. However, the UV divergences cancel in their sum. Therefore it is useful to subtract the UV divergent part of the real correction and add it to the dipole term. Note that now the division of the NLO contributions between the two terms is not unique but depends on the chosen UV subtraction scheme. We choose to do this subtraction following the scheme presented in Ref.~\cite{Hanninen:2017ddy} and used in Refs.~\cite{Mantysaari:2021ryb,Beuf:2021srj}. In our case, this means that we write:
\begin{multline}
	\label{eq:UV_subtraction}
	N_{012} \xt_{20}^i \ical^i_{(j)} K_{(D-2)/2} \big(2 m_q z_2|\xt_{20}|\big)  =\\
	\Big\{ N_{012}\xt_{20}^i  \ical^i_{(j)} K_{(D-2)/2}\big(2 m_q z_2|\xt_{20}|\big)-N_{01} \ical^{q \bar q g}_\text{UV} \Big\} +N_{01} \ical^{q \bar q g}_\text{UV}
\end{multline}
where 
\begin{multline}
	\label{eq:I_UV}
	\ical^{q \bar q g}_\text{UV} = \big(m_q z_2|\xt_{20}|\big)^{-D/2+1}\Gamma(D/2-1)^2 \\
	\times \frac{i\mu^{2-D/2}}{8\pi^{D/2}} |\xt_{20}|^{4-D}  \left(\frac{\zeta}{2\pi |\xt_{01}|^2} \right)^{(D-4)/2}K_{(D-4)/2}\left(\zeta\right) e^{-|\xt_{20}|^2/\left(|\xt_{01}|^2 e^{\gamma_E}\right)}.
\end{multline}
Although the subtraction procedure is not unique,  this particular choice has the correct behavior at $\xt_{20} \rightarrow 0$ and results in relatively simple expressions. With the subtraction of Eq.~\eqref{eq:UV_subtraction} we can perform the $\xt_{20}$ and $z_2$ integrals before adding it to the dipole part:
\begin{equation}
	\label{eq:I_UV_integrated}
	\begin{aligned}
	    & \left[	2 \int_{\xt_0 \xt_1 \xt_2} \int \frac{\dd[]{z_0}\dd[]{z_1} \dd[]{z_2}}{(4\pi)^3} 4\pi \delta(z_0+z_1+z_2-1) \Psi_{\gamma^*}^{q \bar q g}\Psi^{q \bar q g*}_V  N_{012} \right]_\textrm{UV subtraction}\\
		=&\int_0^1 \frac{\dd{z'}}{4\pi}\phi^{q \bar q}(\rt = 0,z')  \int \dd[D-2]{\xt_{01}}\dd[D-2]{\xt_{20}} \dd[2]{\bt}\int_{\alpha}^ {1/2} \dd[]{z_2} \\
		&\times N_{01}\sqrt{\frac{N_c}{2}} \frac{2ee_f m_q}{\pi} \frac{\alpha_s C_F}{2\pi} 4\pi\left(\frac{mz_2}{\pi |\xt_{20}|\mu}\right)^{(D-4)/2}\frac{-8im_q}{|\xt_{20}|}\left[ (D-2)z_2^2-2z_2+1 \right] \ical^{q \bar q g}_\text{UV} \\
		=& \int_0^1 \frac{\dd{z'}}{4\pi}\phi^{q \bar q}(\rt = 0,z')  \int \dd[D-2]{\xt_{01}} \dd[2]{\bt}N_{01}\sqrt{\frac{N_c}{2}} \frac{2e e_f m_q}{\pi} \frac{\alpha_s C_F}{2\pi} K_{(D-4)/2}(\zeta) \\
		&\times \left\{\left(4 \ln(2\alpha)+3\right)\left[\frac{2}{D-4}+\ln(\frac{\zeta}{2\pi^2 |\xt_{01}|^4 \mu^2 e^{\gamma_E}})\right] -1\right\}.
	\end{aligned}
\end{equation}
Adding this to the dipole term we get
\begin{multline}
	\label{eq:dipole_UV_subtracted}
		-i\acal_{q \bar q} =2 \int_0^1 \frac{\dd{z'}}{4\pi}\phi^{q \bar q}(\rt = 0,z')\int \dd[2]{\xt_{01}} \dd[2]{\bt} N_{01}(\rt, \bt) \sqrt{\frac{N_c}{2}} \frac{e e_f m_q}{\pi} \Bigg\{ K_0(\zeta)\\
		+ \frac{\alpha_s C_F}{2\pi} \Bigg\{ \widetilde I^T_{\vcal \mcal \scal}\left(\frac{1}{2}, \xt_{01}\right)+\kcal^T
		+K_0(\zeta)\Bigg[\Omega^T_\vcal\left(\gamma;\frac{1}{2}\right) + L\left(\gamma;\frac{1}{2}\right)- \frac{\pi^2}{3} +\frac{5}{2}\\
		+\frac{1}{\alpha}
		-3\ln(\frac{m_q |\xt_{01}|}{2})-3\gamma_E \Bigg]
		\Bigg\} \Bigg\}.
\end{multline}
Note that this expression is now UV finite, allowing us to take the limit $D \rightarrow 4$. The UV subtraction also cancelled the dependence on the scale $\mu$ introduced by dimensional regularization. 

For the real correction, the UV subtracted form is
\begin{equation}
	\label{eq:real_UV_subtracted}
	-i\acal_{q \bar q g} =
	2 \int_0^1 \frac{\dd{z'}}{4\pi}\phi^{q \bar q}(\rt = 0,z')\sqrt{\frac{N_c}{2}} \frac{e e_f m_q  }{\pi} \int \dd[2]{\xt_{01}} \dd[2]{\bt} \dd[2]{\xt_{20}} \int_\alpha^{1/2} \dd[]{z_2}  \frac{\alpha_s C_F}{2\pi}  \kcal_{q \bar q g}
\end{equation}
where
\begin{multline}
		\label{eq:K_qqg}
		\kcal_{q \bar q g}=
		32\pi m_q \Bigg\{ K_{1}(2 m_q z_2 |\xt_{20}|) \frac{i \xt_{20}^i}{|\xt_{20}|}
		\Bigg[- \ical_{(j)}^i \left((1-z_2)^2+z_2^2 \right) - 
		z_2^2 \left(2z_2-1 \right)\hat \ical_{(j)}^i  \\
		 \shoveright{+\ical_{(k)}^i \frac{1}{2z_2+1}  
		 +\hat \ical_{(k)}^i \frac{z_2^2 (2z_2-1)^2}{(2z_2+1)^2}  \Bigg] N_{012}}\\
		 +\frac{z_2}{m_q} K_{0}\left( 2 m_q z_2|\xt_{20}| \right) \Bigg[ \frac{-1+2z_2}{2}\ical_{(j)}^{ii} 
		 +\frac{1+2z_2}{2}\ical_{(k)}^{ii} 
		 -2 (1-2z_2)z_2 \jcal_{(l)} -4m_q^2 z_2^2 \ical_{(j)}
		 \Bigg]N_{012} \\
		 - \left((1-z_2)^2+z_2^2 \right) \frac{1}{8\pi^2 m_q z_2 |\xt_{20}|^2} K_{0}\left(\zeta\right) e^{-\xt_{20}^2 \B}  N_{01}
		 \Bigg\}.
\end{multline}
This expression is also UV finite and does not depend on the renormalization scale $\mu$.

The dipole term \eqref{eq:dipole_UV_subtracted} still has an apparent IR divergence coming from the $\frac{1}{\alpha}$ term. This is related to the fact that the LOWF is also divergent and has to be renormalized. This can be seen explicitly in the NLO equation for the leptonic width which in the nonrelativistic limit is given by \cite{Escobedo:2019bxn}
\begin{equation}
    \label{eq:Gamma_ee}
    \Gamma(V \rightarrow e^- e^+) = \frac{2N_c e_f^2 e^4}{3 \pi M_V}  \left| \int \frac{\dd[]{z'}}{4 \pi} \phi^{q \bar q}(\rt=0,z')\right|^2 \left[ 1 + \frac{ \alpha_s C_F}{\pi} \left( \frac{1}{\alpha}-4 \right) \right].
\end{equation}
As the leptonic width has to be finite, the LOWF has to have an IR divergent part that cancels the $\frac{1}{\alpha}$ divergence appearing in this equation. One way to account for this IR divergence of the LOWF is to invert Eq.~\eqref{eq:Gamma_ee} and solve the integrated LOWF directly from it, which gives at the order $\mathcal{O}(\as)$
\begin{equation}
    \label{eq:LOWF_decay_width_scheme}
     \int \frac{\dd[]{z'}}{4 \pi} \phi^{q \bar q}(\rt=0, z') = \sqrt{ \Gamma(V \rightarrow e^- e^+) \frac{3 \pi M_V}{2e_f^2 e^4 N_c} }  \left[ 1 + \frac{ \alpha_s C_F}{2\pi} \left( 4-\frac{1}{\alpha} \right) \right].
\end{equation}
This expression can then used in Eqs.~\eqref{eq:dipole_UV_subtracted} and \eqref{eq:real_UV_subtracted} to cancel the IR divergence in the dipole term and to connect the nonperturbative integral over LOWF to the leptonic width for which one can use the experimental value in numerical calculations. The dipole part of the amplitude is then
\begin{equation}
	\label{eq:dipole_correction_dw}
		-i\acal_{q \bar q}	=  \sqrt{ \Gamma(V \rightarrow e^- e^+) \frac{3 \pi M_V}{2N_c e_f^2 e^4} }
		\sqrt{\frac{N_c}{2}}\frac{e e_f m_q}{\pi } 2\int \dd[2]{\xt_{01}}\int \dd[2]{\bt}
		  \left\{ \kcal_{q \bar q}^\lo
		+\frac{\alpha_s C_F}{2\pi}  \kcal_{q \bar q, \Gamma}^\nlo \right\}
\end{equation}
where the LO part is
\begin{equation}
	\label{eq:K_qq_LO}
	\kcal_{q \bar q}^\lo = K_0(\zeta) N_{01},
\end{equation}
and the NLO part, which contains the corrections from virtual gluon loops, is defined as
\begin{multline}
	\label{eq:K_qq_NLO_dw}
	\kcal_{q \bar q, \Gamma}^\nlo = \Bigg\{ \widetilde I^T_{\vcal \mcal \scal}\left(\frac{1}{2}, \xt_{01}\right)+\kcal^T
	\\
	+K_0(\zeta)\left[\Omega^T_\vcal\left(\gamma;\frac{1}{2}\right) + L\left(\gamma;\frac{1}{2}\right)- \frac{\pi^2}{3} +\frac{5}{2}
+4	-3\ln(\frac{m_q|\xt_{01}|}{2})-3\gamma_E \right]\Bigg\} N_{01}.
\end{multline}
We define this way of renormalizing the LOWF to be the \emph{decay width} scheme. Note that this renormalization scheme adds the term $ \frac{\alpha_s C_F}{2\pi} \times 4$ from the equation of the leptonic width to the virtual correction $\kcal_{q \bar q, \Gamma}^\nlo$. 

We can also renormalize the LOWF in a different way where such an additional term does not appear. This is done by connecting the LOWF $\phi^{q\bar q}$ in our regularization scheme to the dimensionally regularized one $\phi^{q\bar q}_\textrm{DR}$ following Ref.~\cite{Escobedo:2019bxn}:
\begin{equation}
	\label{eq:LOWF_wave_function_scheme}
	\int \frac{\dd[]{z'}}{4 \pi} \phi^{q \bar q} = \int \frac{\dd[]{z'}}{4 \pi} \phi^{q \bar q}_\textrm{DR} \times \left[ 1- \frac{\alpha_s C_F}{ 2\pi} \frac{1}{\alpha} \right].
\end{equation}
This also cancels the IR divergence in the virtual correction. Now the dipole  part of the amplitude can be written as
\begin{equation}
	\label{eq:dipole_term_wf}
		-i\acal_{q \bar q}	=  \int \frac{\dd[]{z'}}{4 \pi} \phi^{q \bar q}_\textrm{DR} \times
		\sqrt{\frac{N_c}{2}}\frac{e e_f m_q}{\pi } 2\int \dd[2]{\xt_{01}}\int \dd[2]{\bt}
		  \left\{ \kcal_{q \bar q}^\lo
		+\frac{\alpha_s C_F}{2\pi}  \kcal_{q \bar q, \Psi}^\nlo \right\}
\end{equation}
where the LO part is still given by Eq.~\eqref{eq:K_qq_LO} but the NLO correction is slightly different:
\begin{multline}
	\label{eq:K_qq_NLO_wf}
	\kcal_{q \bar q, \Psi}^\nlo  = \Bigg\{\widetilde I^T_{\vcal \mcal \scal}\left(\frac{1}{2}, \xt_{01}\right)+\kcal^T \\
	+K_0(\zeta)\left[\Omega^T_\vcal\left(\gamma;\frac{1}{2}\right) + L\left(\gamma;\frac{1}{2}\right)- \frac{\pi^2}{3} +\frac{5}{2}
	-3\ln(\frac{m_q|\xt_{01}|}{2})-3\gamma_E \right]\Bigg\} N_{01}.
\end{multline}
We will refer to this renormalization scheme of the LOWF as the \emph{wave function} scheme. Here one still has to determine the value of the dimensionally regularized LOWF for numerical calculations. This can likewise be done with the leptonic width, from which we can solve the dimensionally regularized LOWF in the nonrelativistic limit as
\begin{equation}
\label{eq:dim_reg_LOWF}
     \int \frac{\dd[]{z'}}{4 \pi} \phi^{q \bar q}_\textrm{DR} =  \sqrt{ \Gamma(V \rightarrow e^- e^+) \frac{3 \pi M_V}{2e_f^2 e^4 N_c} }  \left[ 1 + \frac{ \alpha_s C_F}{2\pi} \times 4  \right].
\end{equation}
Note that when the relativistic corrections are taken into account in Sec.~\ref{sec:relativistic} the relation~\eqref{eq:dim_reg_LOWF} will be modified.
The difference between the decay width and wave function schemes boils down to the location of the NLO correction $\frac{\as C_F}{2\pi} \times 4$ from the equation for the leptonic width, i.e., whether the correction appears in the equation for the virtual correction or as an overall coefficient when solving the LOWF. The difference between the schemes is parametrically of the order $\ocal(\as^2)$ which is of higher order than considered here. This difference is also numerically small in realistic kinematics as we will demonstrate in Appendix~\ref{appendix:scheme-dep}.

When choosing the scheme one also has to take into account the running of the coupling constant $\as$. In the decay width scheme we use the running coupling in the coordinate space $\as(\xt_{ij})$, Eq.~\eqref{eq:running_coupling}, whereas in the wave function scheme the coupling constant is evaluated at the momentum scale of the decay process. Following Ref.~\cite{Bodwin:2007fz}, we take this to be the vector meson mass so that the coupling constant in Eq.~\eqref{eq:dim_reg_LOWF} is chosen to be $\as(M_V)$.

\subsection{Rapidity divergence and the leading-order result}
\label{sec:rapidity}

The real correction \eqref{eq:real_UV_subtracted} is still IR divergent as can be verified by taking the $\alpha \rightarrow 0$ limit for the lower bound of the $z_2$ integral. This divergence is actually related to the rapidity evolution of the dipole amplitude. The divergent part of the real correction \eqref{eq:K_qqg} is
\begin{equation}
    \label{eq:K_qqg_BK}
     \frac{\alpha_s C_F}{2\pi}  \kcal_{q \bar q g}^\text{sing} =  \frac{\alpha_s C_F}{\pi^2}  \frac{1}{z_2} K_0(\zeta) \frac{\xt_{01}^2}{\xt_{20}^2 \xt_{21}^2}\left[N_{012}-N_{01}\right]
\end{equation}
from which we recognize, using Eq.~\eqref{eq:S-matrix_012}, the leading-order BK equation \eqref{eq:BK} if the running of the coupling is neglected. Thus at fixed coupling it is possible to combine this divergent part with the leading-order part \eqref{eq:K_qq_LO} of the production amplitude by taking into account the rapidity evolution of the dipole amplitude.

The amount of BK evolution is controlled by the lower limit of the $z_2$ integral (recall that $Y = \ln (z_2 q^+/P^+)$ as discussed in Sec.~\ref{sec:evolution}). 
In practice one should not take here the $\alpha \to 0$ limit. Instead, the lower limit has to be set to a finite value. The reason is the following: as the invariant mass $M_{q \bar q g}$ of the $q \bar q g$ system goes like  $M_{q \bar q g}^2 \sim 1/z_2$ at small $z_2$, in the limit $z_2 \rightarrow 0$ we would have $M_{q \bar q g} \rightarrow \infty$. However, in the calculation we employ the eikonal approximation which assumes that $M_{q \bar q g}^2 \ll W^2$ where $W$ is the center-of-mass energy of the photon-proton system. Therefore we should set the lower limit, denoted by $\zmin$ from now on, such that the eikonal limit is satisfied. We follow Ref.~\cite{Beuf:2020dxl} and choose $\zmin= P^+/q^+$.

The target plus momentum is given by $P^+ = Q_0^2/(2P^-)$ where $Q_0^2$ is again the transverse momentum scale of the target which we have taken to be $Q_0^2= 1\gev^2$ following \cite{Beuf:2020dxl}. The center-of-mass energy of the photon-proton system is then given by $W^2 = 2 q^+ P^- - Q^2 +m_N^2$.
Using these expressions we get the following condition for $z_2$:
\begin{equation}
\label{eq:zmin}
	z_2 > \zmin = \frac{P^+}{q^+} = \frac{Q_0^2}{2P^- q^+} = \frac{Q_0^2}{W^2+Q^2-m_N^2}.
\end{equation}
Using this integration limit we find that the $z_2$ integral of the $\kcal_{q \bar q g}^\text{sing}$ part corresponds to the evolution of the dipole amplitude from the initial rapidity $Y_0=0$ to the rapidity
\begin{equation}
    \Ydip =  \ln \frac{\frac{1}{2}}{\zmin}= \ln \frac{W^2+Q^2 - m_N^2}{2Q_0^2}
\end{equation}
using the leading order BK equation. 

Evaluating the dipole amplitude at this rapidity corresponds to a resummation of large logarithms $\as \ln W^2$. As parametrically $\as \ln W^2 \sim 1$, the actual leading-order part of the production amplitude corresponds to the term $\kcal_{q\bar q}^\lo (\Ydip)$ where we use the rapidity $\Ydip$ to evaluate the dipole amplitude:
\begin{equation}
\label{eq:lo_sub_wf}
	-i \acal_\lo^T = \int \frac{\dd[]{z'}}{4 \pi}  \phi^{q \bar q}_\textrm{DR} \times
	\sqrt{\frac{N_c}{2}}\frac{e e_f m}{\pi } 2\int \dd[2]{\xt_{01}}\int \dd[2]{\bt}
	  \kcal_{q \bar q}^\lo(\Ydip).
\end{equation}
Note that the IR divergent part \eqref{eq:K_qqg_BK} combined with the lowest order part in \eqref{eq:K_qq_LO} results in Eq.~\eqref{eq:lo_sub_wf} (which corresponds to the \emph{subtracted} scheme of Ref.~\cite{Ducloue:2017ftk}) exactly only at fixed coupling. As discussed in Sec.~\ref{sec:evolution}, in this work we use dipole amplitudes that are evolved using running coupling BK equations that include a resummation of most important higher order corrections to all orders and as such approximate the full next-to-leading order BK equation accurately. Consequently the definition of the leading order scattering amplitude is not unique, but we use the definition~\eqref{eq:lo_sub_wf} as it naturally includes also the parametrically large resummation effects included in the used BK equations.

At next-to-leading order, we choose to evaluate the leading-order part at the initial rapidity $Y_0$ and let the $q \bar q g$ part take care of the rapidity evolution. This corresponds to the \emph{unsubtracted} scheme used also in Ref.~\cite{Beuf:2020dxl}. For the virtual correction $\kcal_{q\bar q}^\nlo$ we choose to evaluate the dipole amplitudes at the evolved rapidity $\Ydip$ which corresponds to the total evolution range as discussed above (but note that the dependence on the evolution rapidity is formally of higher order in $\as$).
For the term $\kcal_{q \bar q g}$ in Eq.~\eqref{eq:K_qqg} we use the definition $Y=\ln k_2^+/P^+$ and evaluate the dipole scattering amplitude at the rapidity:
\begin{equation}
	\label{eq:Y_qqg}
	\Yqqg = \ln \frac{z_2 q^+}{P^+} = \ln z_2 + \ln \frac{W^2+Q^2-m_N^2}{Q_0^2}
\end{equation}

Taking the rapidity dependence of the dipole amplitudes into account we can write the final scattering amplitude for the next-to-leading order production amplitude of transversely polarized vector meson in the form
\begin{multline}
	\label{eq:total_NLO_wf}
		-i \acal^T =   \int \frac{\dd[]{z'}}{4 \pi}  \phi^{q \bar q}_\textrm{DR}  \times
		\sqrt{\frac{N_c}{2}}\frac{e e_f m_q }{\pi } 2\int \dd[2]{\xt_{01}}\int \dd[2]{\bt}
		  \Bigg\{ \kcal_{q \bar q}^\lo(Y_0) \\
		+\frac{\alpha_s C_F}{2\pi}  \kcal_{q \bar q, \Psi}^\nlo(\Ydip) 
		+ \int \dd[2]{\xt_{20}} \int_{\zmin}^{1/2} \dd[]{z_2} \frac{\alpha_s C_F}{2\pi} \kcal_{q\bar q g}(\Yqqg)\Bigg\}.
\end{multline}
Here $\kcal_{q \bar q}^\lo$, $\kcal_{q \bar q, \Psi}^\nlo$ and $\kcal_{q\bar q g}$ are given by Eqs.~\eqref{eq:K_qq_LO}, \eqref{eq:K_qq_NLO_wf} and \eqref{eq:K_qqg}, and the rapidity values in parentheses correspond to the rapidities at which the dipole amplitudes are evaluated. An analogous result can also be written in the decay width scheme. The initial rapidity is chosen to be $Y_0=0$ following Ref.~\cite{Beuf:2020dxl}. The strong coupling constant is evaluated at the distance scale set by the parent dipole, $|\xt_{01}|^2,$ in the first two terms and by the smallest dipole $\min\{|\xt_{01}|^2, |\xt_{20}|^2, |\xt_{21}|^2 \}$ in the last term, consistently with the NLO fit of Ref.~\cite{Beuf:2020dxl}.

We note that in Ref.~\cite{Beuf:2020dxl} the virtual contribution $\kcal_{q\bar q}^\nlo$ is evaluated at the rapidity $Y=\ln 1/ \xbj$ which in our case would correspond to $Y^\text{incl}=\ln 1/\xpom \neq \Ydip$.
It is not entirely consistent to use a different evolution rapidity in our NLO calculation of vector meson production and in the fit procedure used to determine the dipole-proton amplitude. However, as in Ref.~\cite{Mantysaari:2021ryb} we choose to use the more natural choice for the evolution rapidity and note that the difference between the usage of the two  rapidities is formally of higher order in $\as$.

\section{Relativistic corrections at leading order}
\label{sec:relativistic}

We can use the nonrelativistic expansion \eqref{eq:NR_expansion} to include the first relativistic corrections of order $v^2$. At leading order in $\as$, the nonrelativistic expansion reduces to a distributional identity
\begin{multline}
	\label{eq:NR_expansion_alphas0}
	\Psi^{V \rightarrow q \bar q}_\lo  =\phi_{h_0 h_1}^{q \bar q}(\rt, z) = \sum_{k} \frac{1}{k_1!k_2!k_3!}  (m_q r_1)^{k_1} (m_q r_2)^{k_2} 4\pi\left(-\frac{1}{2i}\partial_z\right)^{k_3}\delta\left(z-1/2\right) \\
	\times \int_0^1 \frac{\dd{z'}}{4\pi} \frac{1}{m_q^{k_1+k_2}} \partial_{r_1}^{k_1} \partial_{r_2}^{k_2} \phi_{h_0 h_1}^{q \bar q}(\rt = 0, z') [2i(z'-1/2)]^{k_3}
\end{multline}
from which it is easy to read off the coefficients $C^k_{q \bar q \leftarrow q \bar q}$. In general, including terms of order $v^2$ corresponds to including the terms with $|k|=k_1+k_2+k_3 \leq 2$.
From now on, we will focus on the case where the meson polarization is  $\lambda_V=+1$. The final result, Eq.~\eqref{eq:rel_corrections_NRQCD}, will be the same for both transverse polarizations $\lambda_V = \pm 1$. We can then write the relativistic correction at the order $v^2$ as
\begin{multline}
	\label{eq:rel_corrections}
		-i\acal^{\lambda_V=+1}_\text{rel} = \sqrt{\frac{N_c}{2}}\frac{e e_f m_q}{\pi } 2\int \dd[2]{\xt_{01}}\int \dd[2]{\bt} N_{01}\\
		 \times \Bigg\{ K_0(\zeta) \frac{m_q^2}{2} \left[r_1^2 \phi^{q \bar q}_{++}(2,0,0) +r_2^2 \phi^{q \bar q}_{++}(0,2,0) \right] - \frac{r^2 Q^2}{8\zeta} K_1(\zeta) \phi^{q \bar q}_{++}(0,0,2) \\
		-K_1(\zeta) \frac{i\zeta (r_1 - i r_2)}{2r^2} \Big[ r_1 \left( \phi^{q \bar q}_{+-}(1,0,0)-\phi^{q \bar q}_{-+}(1,0,0)  \right) \\
		+ r_2 \left( \phi^{q \bar q}_{+-}(0,1,0) -\phi^{q \bar q}_{-+}(0,1,0) \right) \Big]  \Bigg\}
\end{multline}
where we introduced a simplifying notation
\begin{equation}
	\label{eq:LFWF corrections}
	\phi^{q \bar q}_{h_0 h_1 }(k_1, k_2, k_3) \coloneqq  \int_0^1 \frac{\dd{z'}}{4\pi} \frac{1}{m_q^{k_1+k_2}} \partial_{r_1}^{k_1} \partial_{r_2}^{k_2} \phi_{h_0 h_1}^{q \bar q}(\rt = 0, z') [2i(z'-1/2)]^{k_3}
\end{equation}
and $r_i = (\xt_{01})_i$.
Strictly speaking, it is the complex conjugate of Eq.~\eqref{eq:rel_corrections} that corresponds to vector meson production, but as this quantity is real we have chosen to write it in this form to get rid of the complex conjugates on the meson wave function.
Note that we do not have here terms like $\phi^{q \bar q}_{+-}(1,0,1)$ or $\phi^{q \bar q}_{--}(2,0,0)$. The reason for this is that non-dominant spin components also bring additional velocity suppression, giving a total velocity suppression of $v^{|k|+|\frac{1}{2}(h_0'+h_1')-\lambda_V|}$. This is because in momentum space the meson wave function must have angle-dependence given by $(p_x \pm i p_y)^{|\frac{1}{2}(h_0'+h_1')-\lambda_V|}$ which then has to be coupled with similar terms when combined with the photon wave function. This explains why the non-dominant spin terms have to go like $p_T^{2|\frac{1}{2}(h_0'+h_1')-\lambda_V|} \sim v^{|h_0'+h_1'-2\lambda_V|}$, meaning that there is an additional velocity suppression of $v^{|\frac{1}{2}(h_0'+h_1')-\lambda_V|}$. This can be seen explicitly in Ref.~\cite{Lappi:2020ufv} where $\phi^{\lambda_V=+1}_{+-}(1,0,0) \sim v^2$ and $\phi^{\lambda_V=+1}_{--}(2,0,0)=0$ at $\ocal(v^2)$. 

We can simplify Eq.~\eqref{eq:rel_corrections} by using the spin-parity $J^{PC}=1^{--}$ of the vector meson. Although parity is only a dynamical symmetry in the light-front, one can still use it to derive symmetry relations of the light-front wave function using similar properties such as the so-called mirror parity (see the discussion in Ref.~\cite{Li:2017mlw} and the references therein). The spin of the meson allows us to write the dependence on the azimuthal angle $\varphi_\perp$ as $\phi^{\lambda_V}_{h_0 h_1}(\rt, z)=e^{i m_l \varphi_\perp } \overline \phi^{\lambda_V}_{h_0 h_1}(|\rt|, z)$ where $m_l = \lambda_V- (h_0+h_1)/2$ is the magnetic quantum number and the part $\overline \phi$ does not depend on the angle $\varphi_\perp$. The $C$- and $P$-parities give the requirements $\phi^{\lambda_V}_{h_0 h_1}(\rt, z) = C (-1)^{1+m_l} \phi^{\lambda_V}_{h_1 h_0}(\rt, 1-z)$ and $\phi^{\lambda_V}_{h_0 h_1}(\rt, z) = P (-1)^{m_l+J} e^{i m_l \varphi_\perp} \overline \phi^{-\lambda_V}_{-h_0,-h_1}(|\rt|, z)$. Using these, the relativistic corrections can be simplified as
\begin{multline}
	\label{eq:rel_corrections2}
		-i\acal^{\lambda_V=+1}_\text{rel} = \sqrt{\frac{N_c}{2}}\frac{e e_f m_q}{\pi } 2\int \dd[2]{\xt_{01}}\int \dd[2]{\bt} N_{01}\\
		 \times \Bigg\{\frac{1}{2}m^2 \xt_{01}^2 K_0(\zeta)  \phi^{q \bar q}_{++}(2,0,0)- \frac{r^2 Q^2}{8\zeta} K_1(\zeta) \phi^{q \bar q}_{++}(0,0,2)
		- \zeta K_1(\zeta) \left[i \phi^{q \bar q}_{+-}(1,0,0)\right] \Bigg\}.
\end{multline}

In Ref.~\cite{Mantysaari:2021ryb} where longitudinal production is calculated at NLO, the decay width scheme is used throughout the paper to renormalize the LOWF. This is possible for the longitudinal production even with the relativistic corrections, as it turns out that the leptonic width for a longitudinally polarized vector meson depends only on the fully nonrelativistic part of the LOWF. For transverse production it is no longer possible to use the decay width scheme with the relativistic corrections as then the leptonic width (at leading order in $\as$) has the form
\begin{multline}
    \label{eq:Gamma_ee_rel}
    \Gamma \big(V(\lambda_V = +1) \rightarrow e^- e^+\big)
    = \frac{2N_c e_f^2 e^4}{3\pi M_V}
     \Bigg[\int\frac{\dd{z'}}{4\pi} \frac{1}{2M_V z'(1-z')} \\
    \times  \Big\{m_q \phi_{++}^{q\bar q}(\rt, z')+i(\partial_{\rt_1}-i\partial_{\rt_2}) \left(-z'\phi_{+-}^{q\bar q}(\rt,z')+(1-z')\phi_{-+}^{q\bar q}(\rt,z')\right) \Big\}_{\rt=0}  \Bigg]^2
\end{multline}
which also has contributions from relativistic components of the wave function. The leptonic width~\eqref{eq:Gamma_ee_rel} can be calculated using the light-cone perturbation theory (see e.g. Ref.~\cite{Dosch:1996ss}), and Eq.~\eqref{eq:Gamma_ee_rel} is the form one gets without making any assumptions about the meson wave function. In the nonrelativistic limit $\phi^{q \bar q}_{++} \sim \delta(z'-\frac{1}{2})$, $M_V =2 m_q$, Eq.~\eqref{eq:Gamma_ee_rel} reduces to Eq.~\eqref{eq:Gamma_ee} at LO.

For consistency, one should use the same scheme when combining NLO results of transverse and longitudinal production. We choose to use the wave function scheme throughout this paper as then it is possible to quantify the significance of the relativistic corrections without additional complications coming from the scheme dependence. The previously derived longitudinal cross section  is presented in the wave function scheme in Appendix~\ref{appendix:longitudinal}.

Using Eqs.~\eqref{eq:total_NLO_wf} and \eqref{eq:rel_corrections2} requires that we know the nonperturbative constants related to the LOWF. Note that when relativistic corrections are included we cannot use Eq.~\eqref{eq:LOWF_wave_function_scheme} to directly express dimensionally regularized wave function in terms of the leptonic decay width, as in the case of transverse polarization the relativistic corrections contribute to the decay width as can be seen from Eq.~\eqref{eq:Gamma_ee_rel}. We instead calculate these constants using the heavy vector meson wave function from Ref.~\cite{Lappi:2020ufv} that includes relativistic corrections of order $v^2$. This is a convenient choice as this wave function connects the nonperturbative constants in Eq.~\eqref{eq:total_NLO_wf} and \eqref{eq:rel_corrections2} to the universal NRQCD matrix elements at order $v^2$. Using this wave function allows us to write
\begin{equation}
   \phi^{q \bar q}_\textrm{DR}=\phi^{q \bar q}_{++}(0,0,0) = \frac{1}{\sqrt{4 m_q}} \left[\phi_\text{RF}(0) + \frac{7}{12m_q^2}\vec \nabla^2 \phi_\text{RF}(0)\right],
\end{equation}
\begin{equation}
    \phi^{q \bar q}_{++}(2,0,0) = \phi^{q \bar q}_{++}(0,0,2) = 2[i\phi^{q \bar q}_{+-}(1,0,0)] = \frac{1}{6\sqrt{m_q}} \frac{1}{m_q^2}\vec \nabla^2 \phi_\text{RF}(0)
\end{equation}
where $\phi_\text{RF}(\vec r)$ is the value of the \emph{rest-frame} wave function in the position space (see Ref.~\cite{Lappi:2020ufv} for the corresponding light-front wave function). One advantage of using this particular choice of the wave function is that it results in the same leptonic width for both the longitudinal and transverse polarizations at the order $\ocal(v^2)$, which follows from the spherical symmetry of the wave function in the rest frame.

The rest-frame wave function can be connected to the NRQCD matrix elements by~\cite{Bodwin:2007fz}
\begin{equation}
	\label{eq:RFWF_v0}
	\phi_\text{RF}( 0) = \frac{1}{\sqrt{2N_c}} \sqrt{\OLDME} \left[ 1+ \ocal\left(v^4\right) \right],
\end{equation}
\begin{equation}
	\label{eq:RFWF_v2}
	\vec \nabla^2 \phi_\text{RF}(0)  = -\qLDME \phi_\text{RF}( 0) \left[ 1+ \ocal\left(v^2\right) \right]
	=-\qLDME \frac{1}{\sqrt{2N_c}} \sqrt{\OLDME}\left[ 1+ \ocal\left(v^2\right) \right].
\end{equation}
Numerical values for the NRQCD matrix elements can be obtained from the decay width data. For \jpsi production, we use the values for  $\OLDME$ and $\qLDME$ (with  their correlated uncertainties) from Ref.~\cite{Bodwin:2007fz}. Similarly, for $\Upsilon$ production the matrix elements from Ref.~\cite{Chung:2010vz} are used. We note that the matrix elements for \jpsi in Ref.~\cite{Bodwin:2007fz} are determined using a charm mass $m_c=1.4\gev$. On the other hand, in our calculation we use the non-relativistic value $m_c= M_V/2$, where $M_V$ is the \jpsi mass, effectively neglecting the quark momentum contribution to the meson invariant mass (similarly the $b$ quark mass is taken to be half of the $\Upsilon$ mass). Consequently, the different mass values result in a difference which is of higher order in quark velocity $v$, see also the discussion in Ref.~\cite{Lappi:2020ufv}.

The relativistic corrections to the production amplitude can then be written as
\begin{multline}
	\label{eq:rel_corrections_NRQCD}
		-i\acal^T_\text{rel} = \sqrt{\frac{N_c}{2}}\frac{e e_f m_q}{\pi } 2\int \dd[2]{\xt_{01}}\int \dd[2]{\bt} N_{01} \\
		\times \frac{1}{6\sqrt{m_q}} \frac{1}{m_q^2} \vec \nabla^2 \phi_\text{RF}(0) \Bigg\{ \frac{1}{2} m_q^2 \xt_{01}^2 K_0(\zeta) - \frac{\xt_{01}^2 Q^2}{8\zeta} K_1(\zeta)-\frac{1}{2} \zeta K_1(\zeta)  \Bigg\}.
\end{multline}
The main result of this work, the exclusive heavy vector meson production cross section at the order $\mathcal{O}(\as v^0, \as^0 v^2)$ is then Eq.~\eqref{eq:total_NLO_wf}, to which Eq~\eqref{eq:rel_corrections_NRQCD} is added.

\section{Numerical results}
\label{sec:numerical}

We show numerical results for the transverse vector meson production amplitude and for the total (longitudinal and transverse) coherent vector meson production cross section in $\gamma^*+p$ scattering at next-to-leading order. We consider separately the fully nonrelativistic limit $\mathcal{O}(\as v^0)$, and the case where first relativistic corrections are included, $\mathcal{O}(\as v^0, \as^0 v^2)$. The longitudinal vector meson production is calculated using the results of Ref.~\cite{Mantysaari:2021ryb} included for completeness in Appendix~\ref{appendix:longitudinal}.

The numerical calculations are done in the wave function scheme for the renormalization of the LOWF. We note that it would be possible to use the decay width scheme in the nonrelativistic limit, but this would introduce additional scheme dependence when comparing to the case with the relativistic corrections included. The difference between the two wave function renormalization schemes is studied numerically in detail in Appendix~\ref{appendix:scheme-dep}.

\subsection{Transverse vector meson production amplitude}

\begin{figure}
	\centering
    \begin{subfigure}{0.49\textwidth}
        \centering
        \includegraphics[width=\textwidth]{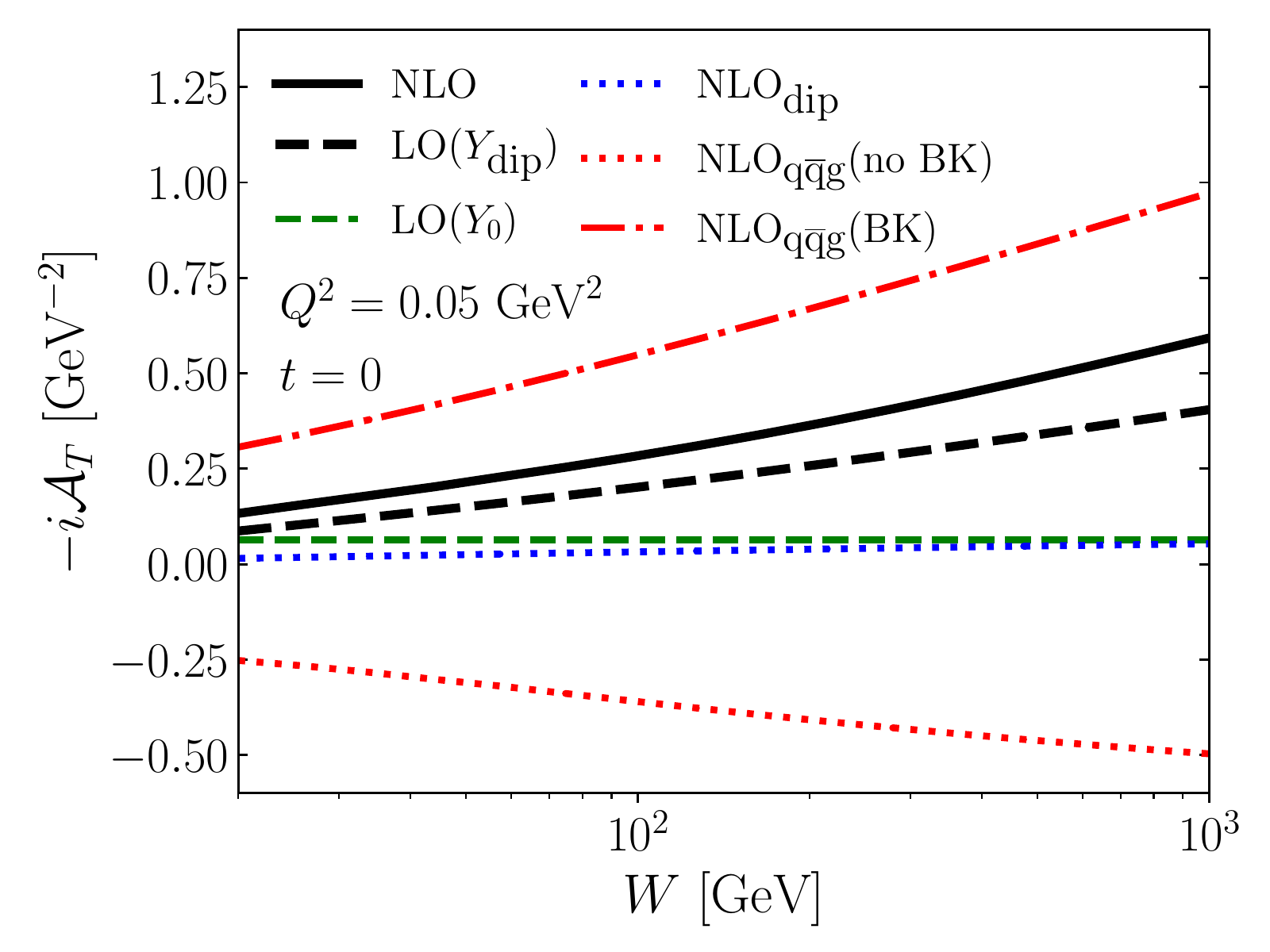}
        \caption{ Amplitude as a function of the center-of-mass energy $W$.}
    \end{subfigure}
    \begin{subfigure}{0.49\textwidth}
        \centering
        \includegraphics[width=\textwidth]{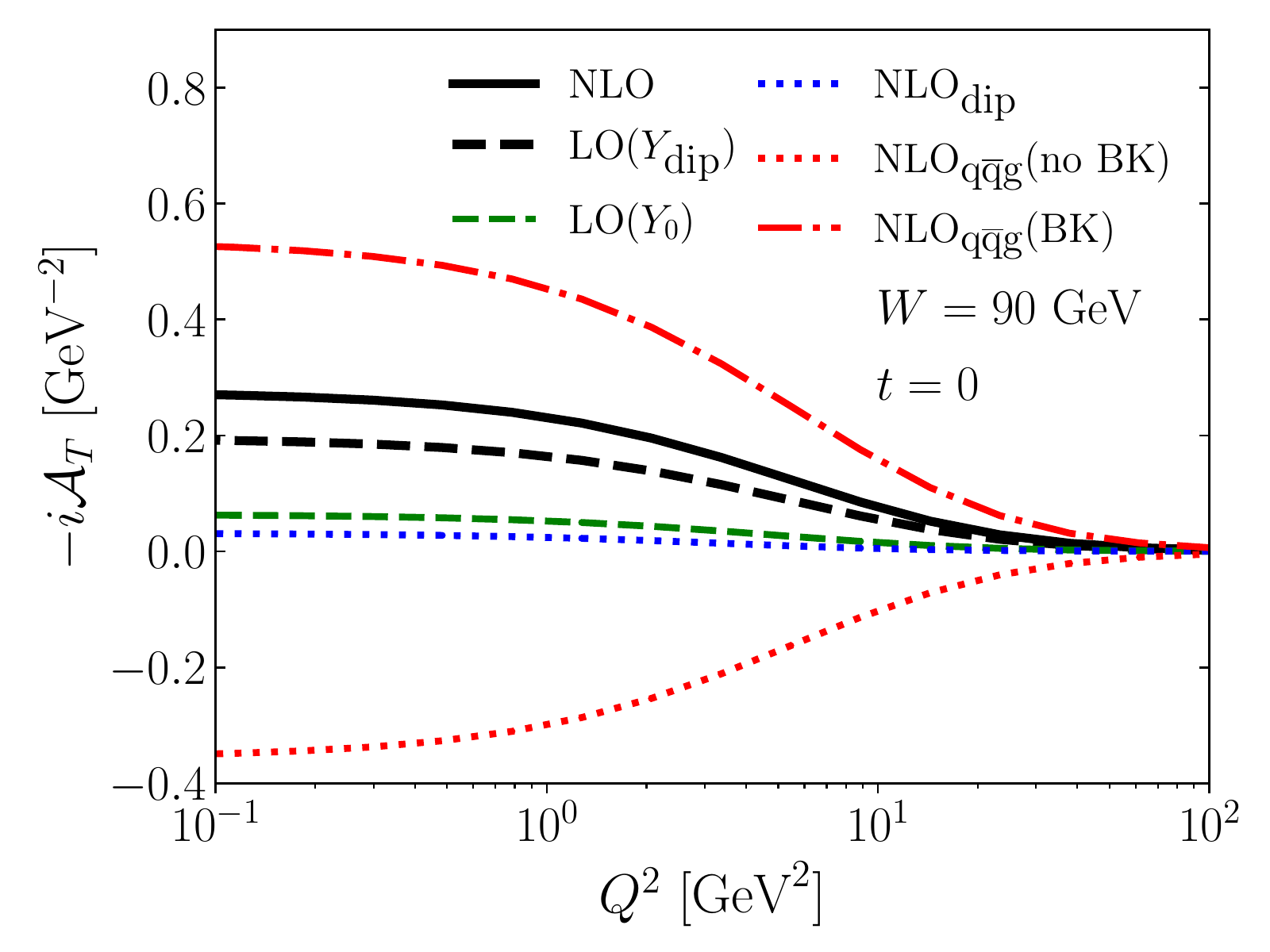}
        \caption{ Amplitude as a function of the photon virtuality $Q^2$.}
    \end{subfigure}
	  \caption{Different contributions to the transverse exclusive \jpsi production scattering amplitude at next-to-leading order. }
        \label{fig:transverse_amplitude}
\end{figure}

First we study in detail different contributions to the forward ($t=0$) transverse \jpsi\ production amplitude, Eq.~\eqref{eq:total_NLO_wf}, in the nonrelativistic limit. This equation is finite and can be directly evaluated numerically. We limit ourselves to the forward production case as we do not want to specify any specific form of impact parameter dependence for the dipole-proton scattering amplitude, and at $t=0$ only the dipole-proton amplitude integrated over the impact parameter appears.  The dipole-proton amplitude is obtained from Ref.~\cite{Beuf:2020dxl} where the assumption is that the impact parameter $\bt$ integral only results in a constant factor $\sigma_0/2$ interpreted as the proton transverse area, also determined from the fit to HERA structure function data.

Different contributions to the scattering amplitude as a function of center-of-mass energy $W$ and photon virtuality $Q^2$ are shown in Fig.~\ref{fig:transverse_amplitude}. In these calculations we have chosen to use as the dipole amplitude the fit performed with the KCBK evolution and using an initial evolution rapidity $Y_{0,\textrm{BK}}=4.61$ in Ref.~\cite{Beuf:2020dxl}. Note that in Ref.~\cite{Beuf:2020dxl} two different initial rapidities for the BK evolution are used ($Y_{0,\textrm{BK}}=4.61$ and $Y_{0,\textrm{BK}}=0$), and the dipole amplitude is frozen in the region $Y_0 < Y <Y_{0,\textrm{BK}}$. Results using both of these initial evolution rapidities are shown later in this Section.

The different contributions to the scattering amplitude shown in Fig.~\ref{fig:transverse_amplitude} are labeled as follows. First, NLO corresponds to the full NLO level scattering amplitude of Eq.~\eqref{eq:total_NLO_wf}. The leading-order result, obtained using a BK-evolved dipole amplitude evaluated at the rapidity $Y=\Ydip$ is labeled as $\lo(\Ydip)$ and shown in Eq.~\eqref{eq:lo_sub_wf}, and $\lo(Y_0)$ corresponds to the $\sim \as^0$ part (first line of Eq.~\eqref{eq:total_NLO_wf}) where the dipole amplitude is evaluated at the initial rapidity. The virtual NLO contribution is denoted by $\nlodip$, but we emphasize that an UV divergence has been cancelled between the real and virtual contributions and as such the division of NLO corrections into the real and virtual parts is not unique. The real gluon emission correction obtained using the UV subtraction scheme used in this work, Eq.~\eqref{eq:real_UV_subtracted}, is shown as $\nloqqg(\textrm{BK})$ and $\nloqqg(\textrm{no BK})$. The $\nloqqg(\textrm{BK})$ term refers to the singular part of the gluon emission contribution (see Eq.~\eqref{eq:K_qqg_BK})  which can be included in the BK evolution, and $\nloqqg(\textrm{no BK})$ corresponds to the remaining pure NLO correction. In this notation the total NLO amplitude can be expressed as
\begin{equation}
\nlo = \lo(Y_0) + \nlodip + \nloqqg(\textrm{BK}) + \nloqqg(\textrm{no BK}).
\end{equation}

The real gluon emission contribution ($\nloqqg$ terms) has a large contribution, which is expected as the BK evolution resumming terms $\sim \as \ln 1/x \sim 1$ to all orders should be considered to be part of the leading order result. However, we also find a significant negative NLO correction $\nloqqg(\textrm{no BK})$ to the BK evolution from the actual NLO calculation where the exact gluon emission kinematics is included. In our UV subtraction scheme the virtual NLO contribution $\nlodip$ is very small.
The total NLO correction is significant, about $\sim 50\%$ of the (BK-evolved) LO contribution. These conclusions are valid at all $W$ and $Q^2$.

In the longitudinal production case presented in Ref.~\cite{Mantysaari:2021ryb}, the NLO corrections were found to be even more significant ($\sim 75 \%$ of the LO result); however, these results cannot be directly compared as the longitudinal calculation used the decay width scheme for the renormalization of the LOWF as opposed to the wave function scheme. In the decay width scheme the NLO corrections are larger because of the larger scheme-dependent constant in $\kcal_\nlo^{q \bar q}$, which is true for both longitudinal and transverse production. It should be noted that this difference in the NLO corrections is compensated by the overall LOWF-related constant in the amplitude which is smaller in the decay width scheme, bringing the numerical values of the full NLO result in the two schemes closer to each other and thus reducing the scheme dependence.

The term $\nloqqg(\textrm{BK})$ corresponds to the (leading order) BK evolution, and as such one could also take the leading order scattering amplitude to be $\lo(Y_0)+\nloqqg(\textrm{BK})$. As can be seen in Fig.~\ref{fig:transverse_amplitude}, this differs from $\lo(\Ydip)$ by roughly a factor of $2$. As discussed in Sec.~\ref{sec:rapidity} this difference would vanish at fixed coupling if the dipole amplitude satisfied the leading order BK equation, but in this work where we use resummed BK evolution equations to approximate the full NLO BK equation it is more natural to use $\lo(\Ydip)$ as a leading order amplitude. This choice includes most of the parametrically large resummation corrections to the leading order amplitude and renders the NLO corrections moderate. On the other hand, if one used $\lo(Y_0)+\nloqqg(\textrm{BK})$ as a leading order amplitude, the NLO corrections would be dominant and even render the cross section negative at high $Q^2$.

\subsection{Differential cross section at $t=0$}

\begin{figure}
	\centering
    \begin{subfigure}{0.45\textwidth}
        \centering
        \includegraphics[width=\textwidth]{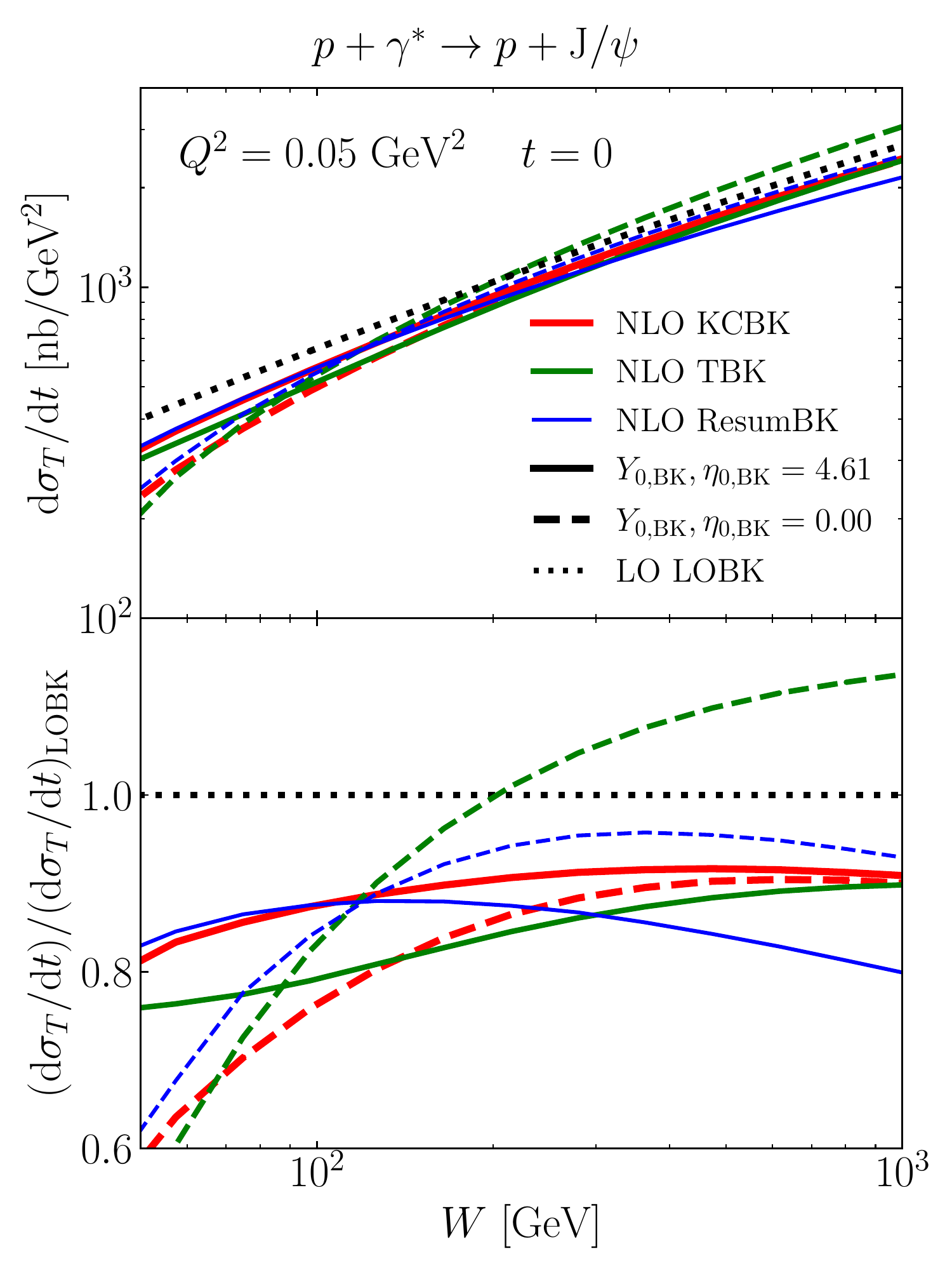}
        \caption{ Comparison of results with different dipole amplitudes. }
        \label{fig:transverse_xs_dipole_amplitude_comparison}
    \end{subfigure}
    \begin{subfigure}{0.45\textwidth}
        \centering
        \includegraphics[width=\textwidth]{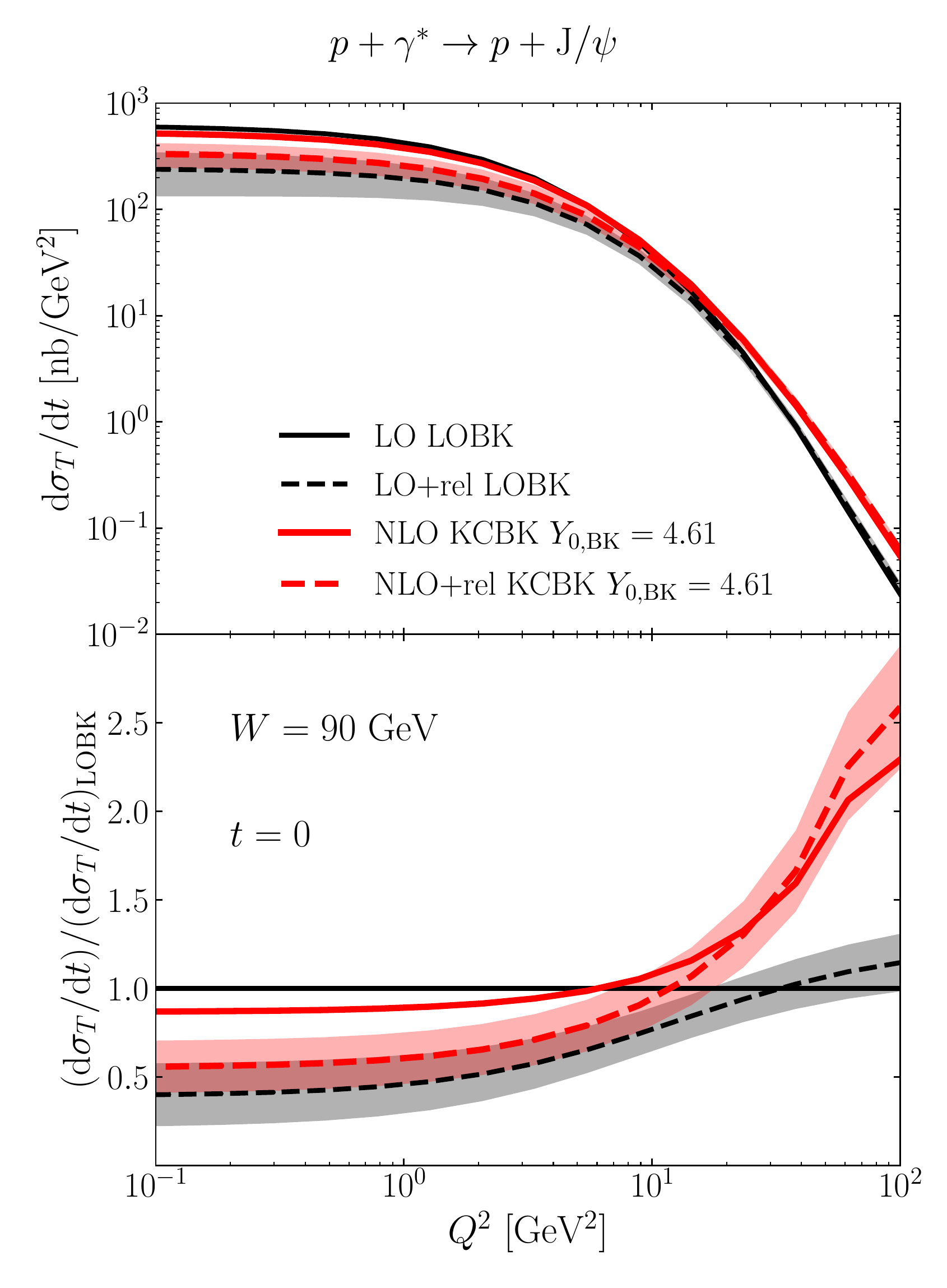}
        \caption{ Comparison of the relativistic and next-to-leading order corrections.  }
        \label{fig:transverse_xs_kcbk}
    \end{subfigure}
	  \caption{Differential cross section for transverse \jpsi production at next-to-leading order.
	  In the case of TBK evolution the rapidity values correspond to the target rapidity $\eta$. The uncertainty bands are obtained by taking into account the uncertainties of the NRQCD matrix elements.
	  }
        \label{fig:transverse_xs}
\end{figure}

Next we calculate the coherent transverse \jpsi production cross section at $t=0$. 
When squaring the scattering amplitude only the genuine NLO corrections are kept, and NNLO contributions proportional to $\as^2$ are dropped out. In practice, Eq.~\eqref{eq:lo_sub_wf} is used as the leading-order amplitude, and when squaring the amplitude its interference with genuine NLO contributions is needed. This NLO correction is obtained from Eq.~\eqref{eq:total_NLO_wf} by subtracting the leading-order amplitude~\eqref{eq:lo_sub_wf}. Similarly, when including relativistic corrections we do not keep the square of the relativistic correction that would be proportional to $v^4$ (but note that such contribution was included in the numerical results reported in Ref.~\cite{Mantysaari:2021ryb}). Instead, when relativistic corrections are included we add the interference term between the leading-order amplitude, Eq.~\eqref{eq:lo_sub_wf}, and the $v^2$ suppressed part of the amplitude, Eq.~\eqref{eq:rel_corrections_NRQCD}.

The differential \jpsi production cross section at $t=0$ is shown in  Fig.~\ref{fig:transverse_xs_dipole_amplitude_comparison} as a function of the center-of-mass energy $W$, using different fits for the dipole-proton scattering amplitude from Ref.~\cite{Beuf:2020dxl}. Results obtained using fits where the initial evolution rapidity is $\Ybk=4.61$ (or $\etabk=4.61$ in the case of TBK evolution) are shown as solid lines, and dashed lines correspond to calculations where fits with the initial rapidity $\Ybk=0$ ($\etabk=0$) are used.
The LO result is ``LO LOBK" which uses the leading-order dipole-proton amplitude from Ref.~\cite{Lappi:2013am} (we use the fit referred to as ``MV$^e$'' in~\cite{Lappi:2013am}) at the rapidity $Y=\ln 1/\xpom$ as this is the rapidity scale used in the LO fit.

We see that the NLO corrections reduce the cross section slightly. This is in contrast to what is seen in Fig.~\ref{fig:transverse_amplitude}, and can be explained by the fact that in Fig.~\ref{fig:transverse_amplitude} the same NLO-fitted dipole amplitude was used for both the LO and NLO results. When nonperturbative parameters describing the initial condition for the BK evolution are determined in leading-order fits such as in Refs.~\cite{Lappi:2013am,Albacete:2010sy}, they effectively absorb a part of the higher-order contributions.

These results are in line with what has been obtained for longitudinal vector meson production at NLO \cite{Mantysaari:2021ryb}. The NLO corrections generally change the center-of-mass energy dependence (faster evolution at low $W$ compared to LO, slower or similar to LO at high $W$). We also find some deviations between the results with different NLO dipole amplitudes, similarly to the longitudinal \jpsi production case~\cite{Mantysaari:2021ryb}. As all of these dipole amplitudes were fitted to the same HERA structure function data, this deviation shows that vector meson production gives us complementary information to structure function analyses. This will be discussed more in Sec.~\ref{sec:total_production}.

In Fig.~\ref{fig:transverse_xs_kcbk}, we show the effect of relativistic corrections at LO and NLO. At small $Q^2$, the relativistic corrections reduce the cross section by $\sim 60\%$ at LO and $\sim 40\%$ at NLO and are numerically more important than the next-to-leading order QCD corrections, that in turn have a much larger effect at large $Q^2$. 
The relativistic corrections have a smaller relative effect at NLO than at LO because of the sizable NLO corrections (recall that we do not include $\mathcal{O}(\as v^2)$ corrections).
It should be noted that the relativistic corrections  do not vanish at high $Q^2$, which is in contrast with longitudinal production where the relativistic corrections are negligible for high photon virtualities \cite{Hoodbhoy:1996zg, Mantysaari:2021ryb}.  

\subsection{Total vector meson production cross section}
\label{sec:total_production}
The transverse vector meson production can be combined with longitudinal production to calculate the total vector meson production cross section $\sigma_\textrm{tot} =\sigma_L + \sigma_T$. This is a phenomenologically more interesting quantity as most exclusive vector meson production data is measured in terms of the total production cross section. 

\begin{figure}
	\centering
    \begin{subfigure}{0.47\textwidth}
        \centering
        \includegraphics[width=\textwidth]{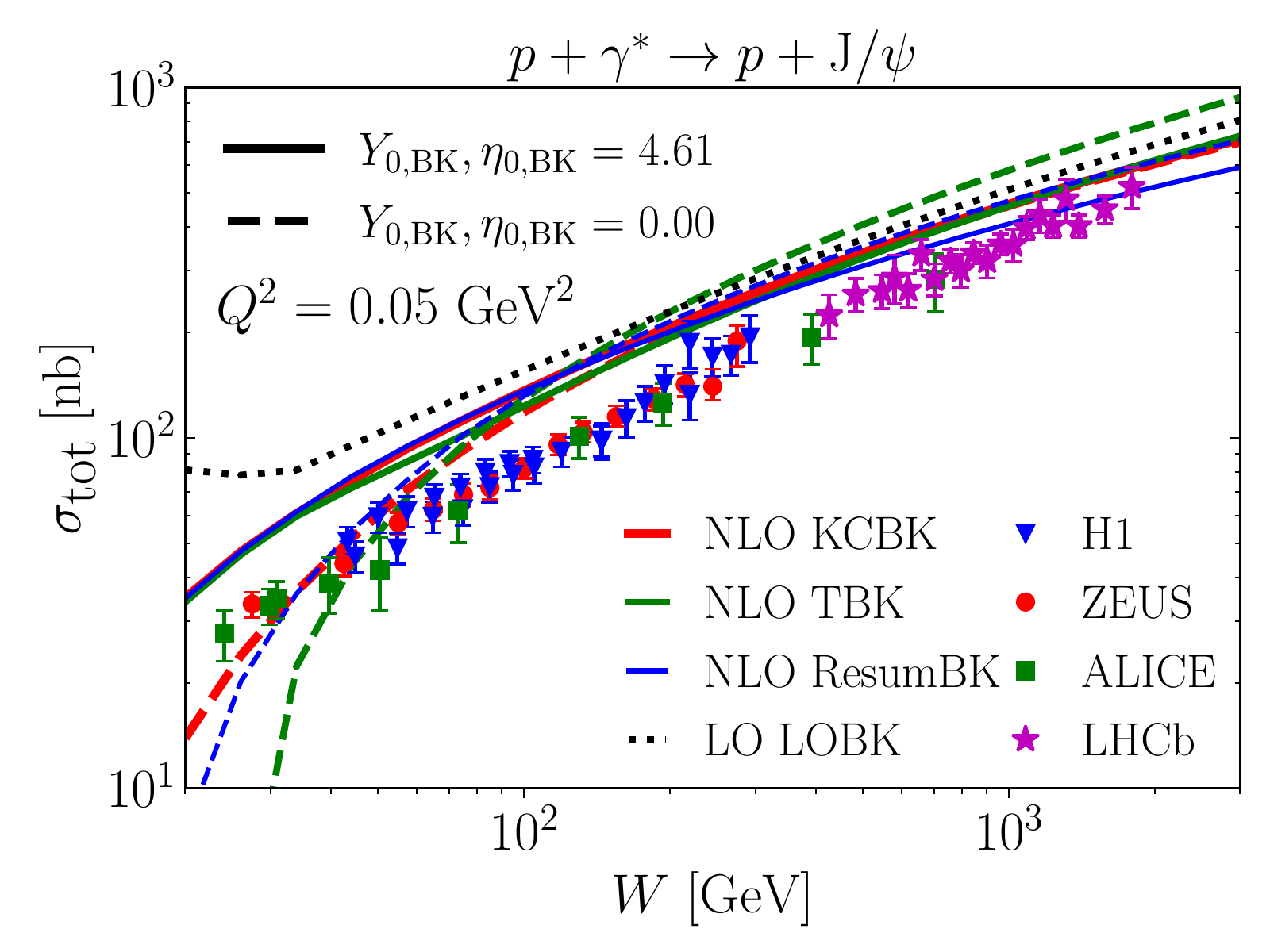}
        \caption{Cross section at the nonrelativistic limit as a function of the center-of-mass energy $W$. 
        }
    \end{subfigure}\quad 
    \begin{subfigure}{0.47\textwidth}
        \centering
        \includegraphics[width=\textwidth]{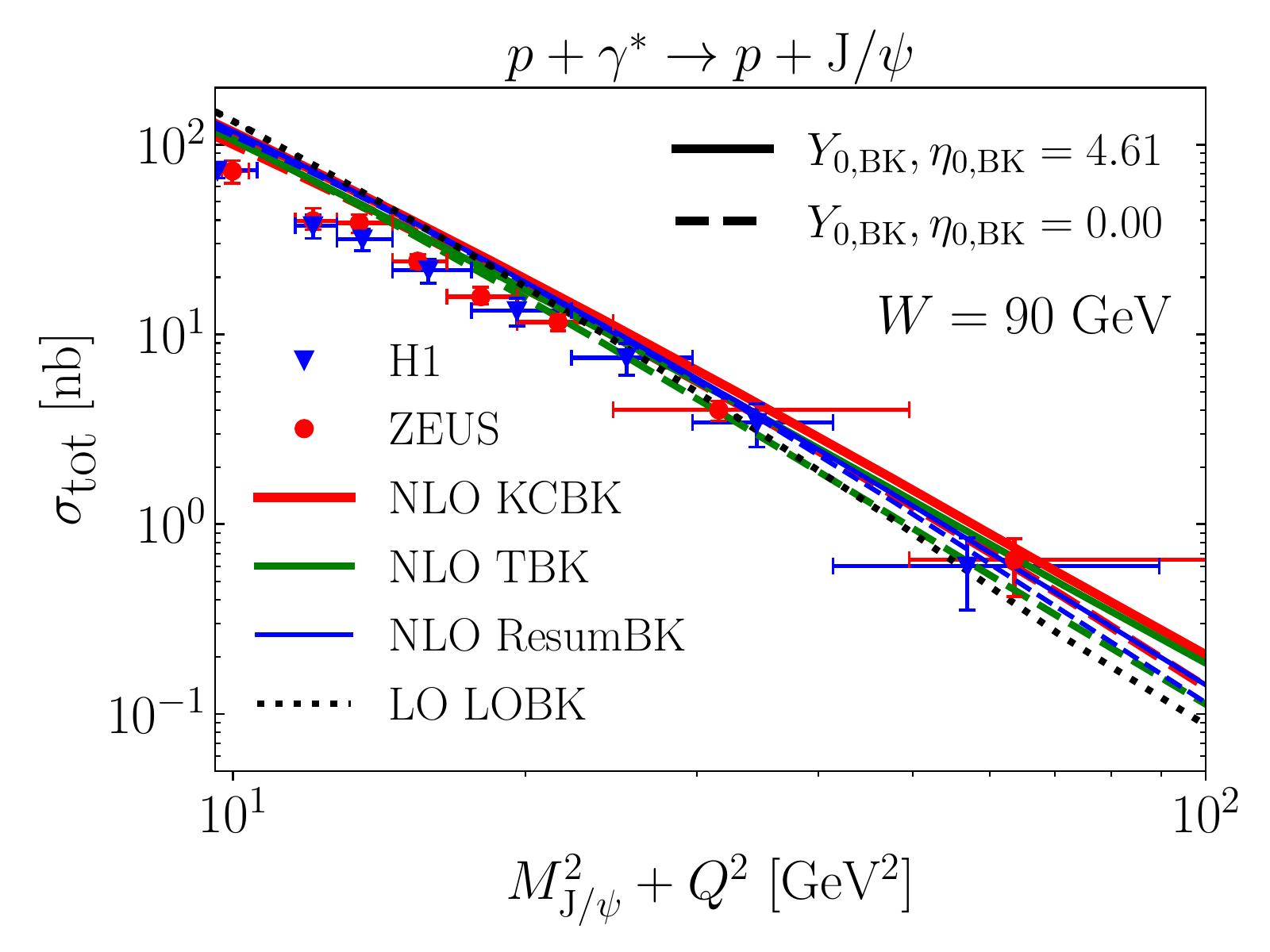}
        \caption{ Cross section at the nonrelativistic limit as a function of the photon virtuality $Q^2$.}
    \end{subfigure}
    \begin{subfigure}{0.47\textwidth}
        \centering
        \includegraphics[width=\textwidth]{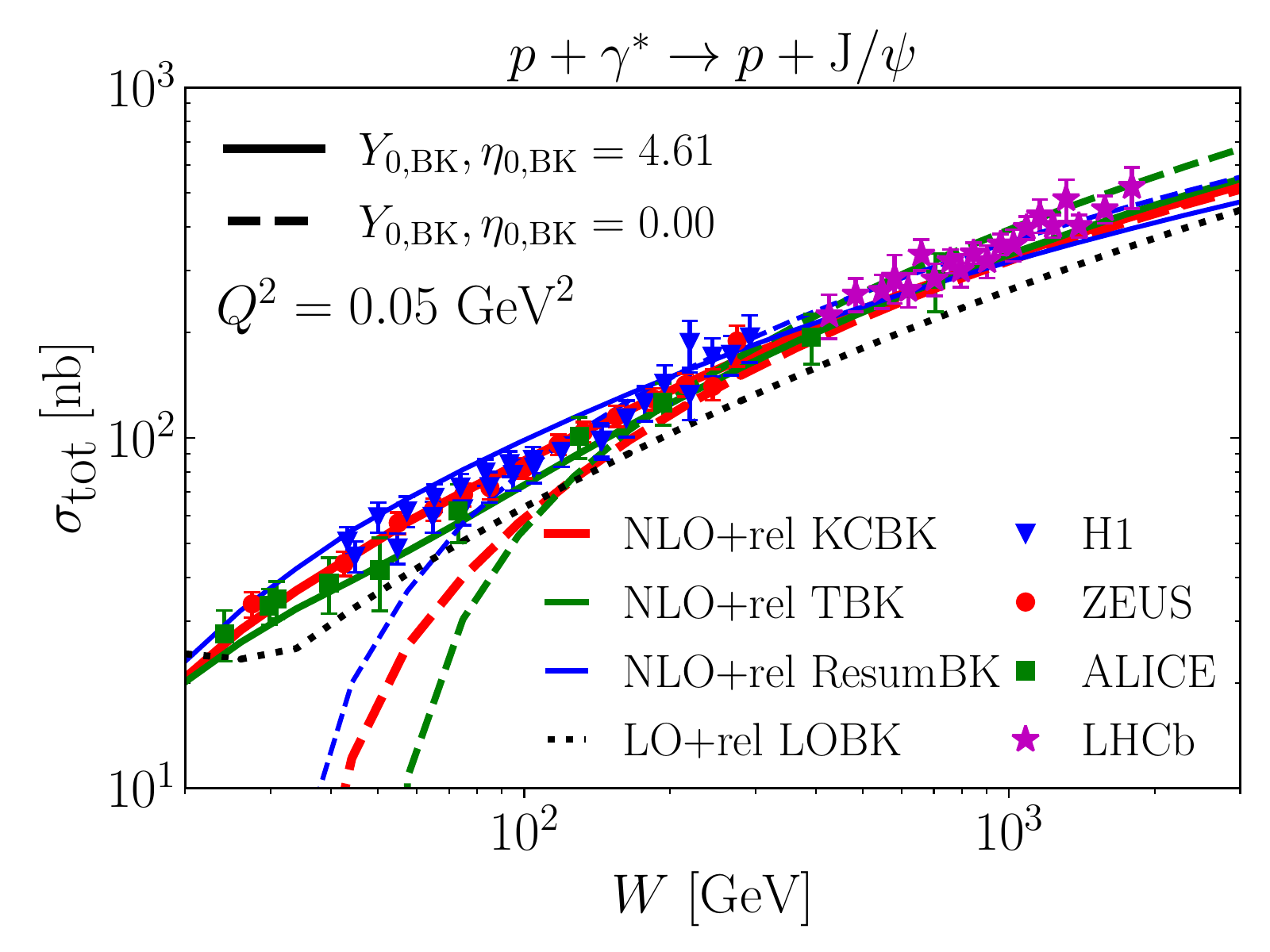}
        \caption{ Cross section with $v^2$ corrections as a function of the center-of-mass energy $W$.}
    \end{subfigure}\quad
    \begin{subfigure}{0.47\textwidth}
        \centering
        \includegraphics[width=\textwidth]{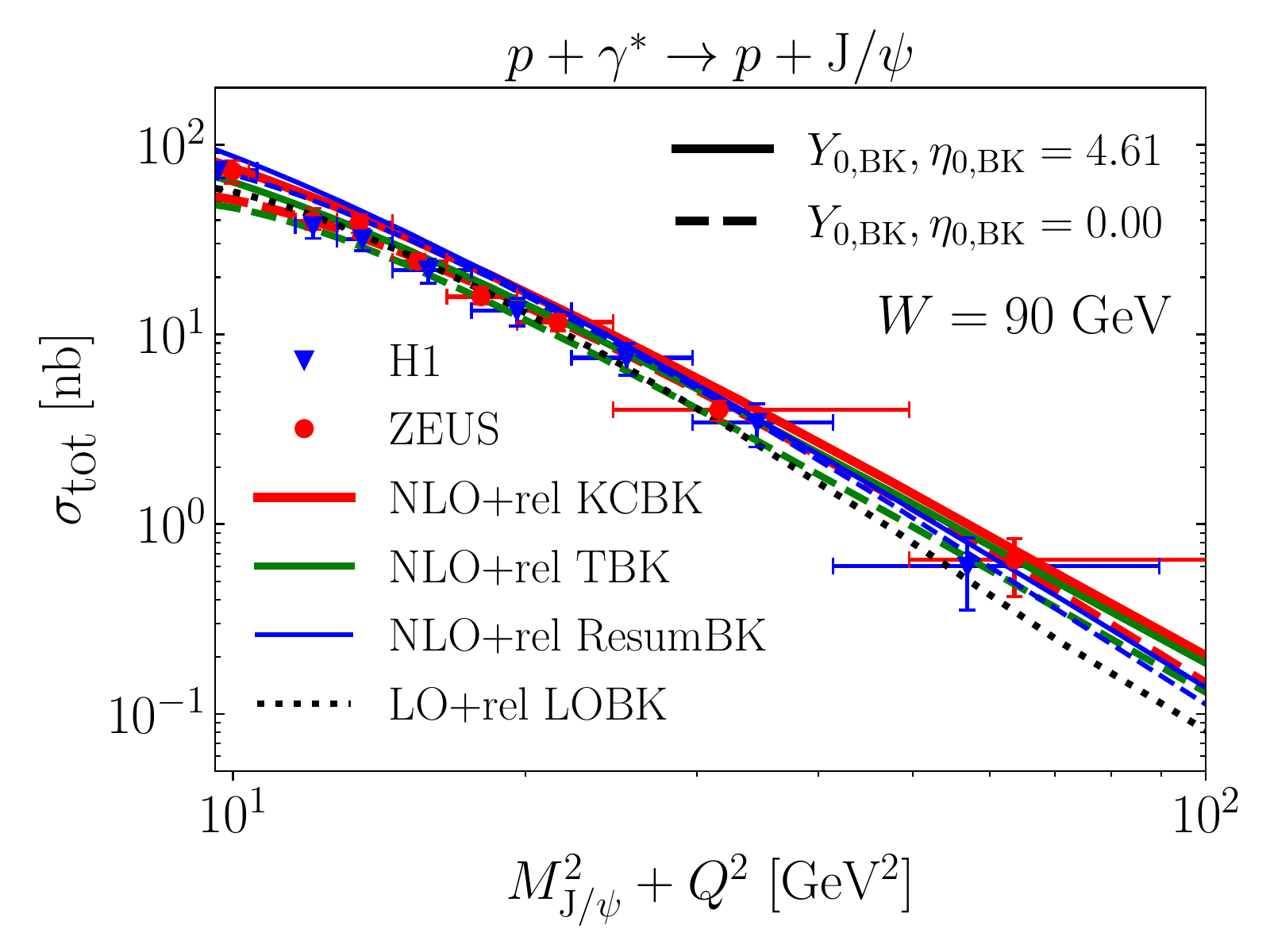}
        \caption{ Cross section with $v^2$ corrections as a function of the photon virtuality $Q^2$.}
    \end{subfigure}
	   \caption{Total exclusive \jpsi production at next-to-leading order, with HERA data from \cite{H1:2005dtp,H1:2013okq,ZEUS:2002wfj,ZEUS:2004yeh}, ALICE data from \cite{ALICE:2014eof,ALICE:2018oyo}, and LHCb data from \cite{LHCb:2014acg, LHCb:2018rcm}.
	   }
        \label{fig:total_xs}
\end{figure}

As discussed above, we only calculate vector meson production at $t=0$ (see Eq.~\eqref{eq:im_amplitude})  as we do not want to specify any particular model for the proton impact parameter profile. In order to obtain results that can be compared with experimental $t$-integrated cross section measurements, we use the experimentally measured $t$ slopes that allow us to write the $\gamma^* + p \to V+p$ cross section as
\begin{equation}
    \label{eq:xs_t-dependence}
    \frac{\dd{\sigma}}{\dd{t}} = e^{-b|t|} \times \frac{\dd{\sigma}}{\dd{t}}(t=0).
\end{equation}
For the \jpsi production the slope $b$ can be written as
$b=b_0 + 4 \alpha^\prime \ln(W/(90\gev))$, where the experimentally measured values for the \jpsi production are $b_0=4.15\gev^{-2}$ and $\alpha^\prime=0.116\gev^{-2}$~\cite{ZEUS:2002wfj}. This allows us to calculate the total $t$-integrated vector meson production. 

The results are shown in Fig.~\ref{fig:total_xs} separately for the nonrelativistic case and with the $v^2$ relativistic corrections, and they are compared to the experimental data measured by H1~\cite{H1:2005dtp,H1:2013okq}, ZEUS~\cite{ZEUS:2002wfj,ZEUS:2004yeh}, ALICE~\cite{ALICE:2014eof,ALICE:2018oyo}, and LHCb~\cite{LHCb:2014acg, LHCb:2018rcm} collaborations. However, we emphasize again that the phenomenological analysis here is not fully consistent as we are using the dipole amplitudes extracted from a fit to structure function data where only the light quark contribution is included~\cite{Beuf:2020dxl}, and a fully consistent setup would require a heavy quark contribution to be included in the structure function calculations also. Consequently, strong conclusions cannot be drawn from these data comparisons.

Keeping this uncertainty in mind, we find that both the $W$ and $Q^2$ dependence of the 
experimental
data is described reasonably well, especially when the relativistic corrections are included. For the virtuality dependence, the relativistic corrections are important at low $Q^2$ and the next-to-leading order corrections also modify the dependence on $Q^2$ slightly, however both LO and NLO results are compatible with the HERA data. Generally, we again find that both the relativistic and NLO corrections can be numerically important and need to be included when considering the \jpsi production.

At $W \lesssim 100 \gev$ the calculations using dipole amplitudes from BK fits where the evolution starts at the smallest possible evolution rapidity $\Ybk=0$ (or $\etabk=0$ in the case of TBK evolution) result in $W$ slope which is not compatible with the data. The next-to-leading order corrections also become extremely large, even rendering the cross section negative. As we will demonstrate in Appendix~\ref{appendix:scheme-dep}, the results obtained with $\Ybk=0$ ($\etabk=0$) in the low-$W$ region are also sensitive to the wave function renormalization scheme, but this is not the case for the fits with $\Ybk=4.61$ ($\etabk=4.61$). We consider this behavior in the low-energy region to be an artifact of the unphysical initial condition obtained in the BK evolution fits in Ref.~\cite{Beuf:2020dxl} when the evolution is started at $\Ybk=0$, in which case there is a long evolution before one enters in the region probed by small-$x$ structure function data. In that case the  fit results in unphysical parameters, and especially the anomalous dimension $\gamma$ is very large at the initial condition\footnote{The dipole amplitude behaves as $N_{01} \sim (\xt_{01}^2 Q_s^2)^\gamma$ in the dilute region, and large $\gamma$ corresponds to e.g. negative unintegrated gluon distribution~\cite{Giraud:2016lgg}.}. As heavy vector meson production is sensitive to smaller size dipoles than the structure function fitted in~\cite{Beuf:2020dxl}, similar unrealistically large NLO corrections were not observed in the NLO fit of Ref.~\cite{Beuf:2020dxl}. Consequently, we consider the results obtained with $\Ybk=4.61$ to be our main numerical results and emphasize that heavy vector meson production data provides additional constraints for the determination of the nonperturbative initial condition for the small-$x$ evolution.

The leading-order result is constant at $W<31\gev$ which corresponds to $\xpom > 0.01$. This is because in the leading-order calculation the dipole amplitude is evaluated at $Y=\ln 1/ \xpom$ and no BK evolution is included in the region $Y<\ln \frac{1}{0.01}$ in the leading order fit~\cite{Lappi:2013zma} used in this work.

\begin{figure}
	\centering
    \begin{subfigure}{0.45\textwidth}
        \centering
        \includegraphics[width=\textwidth]{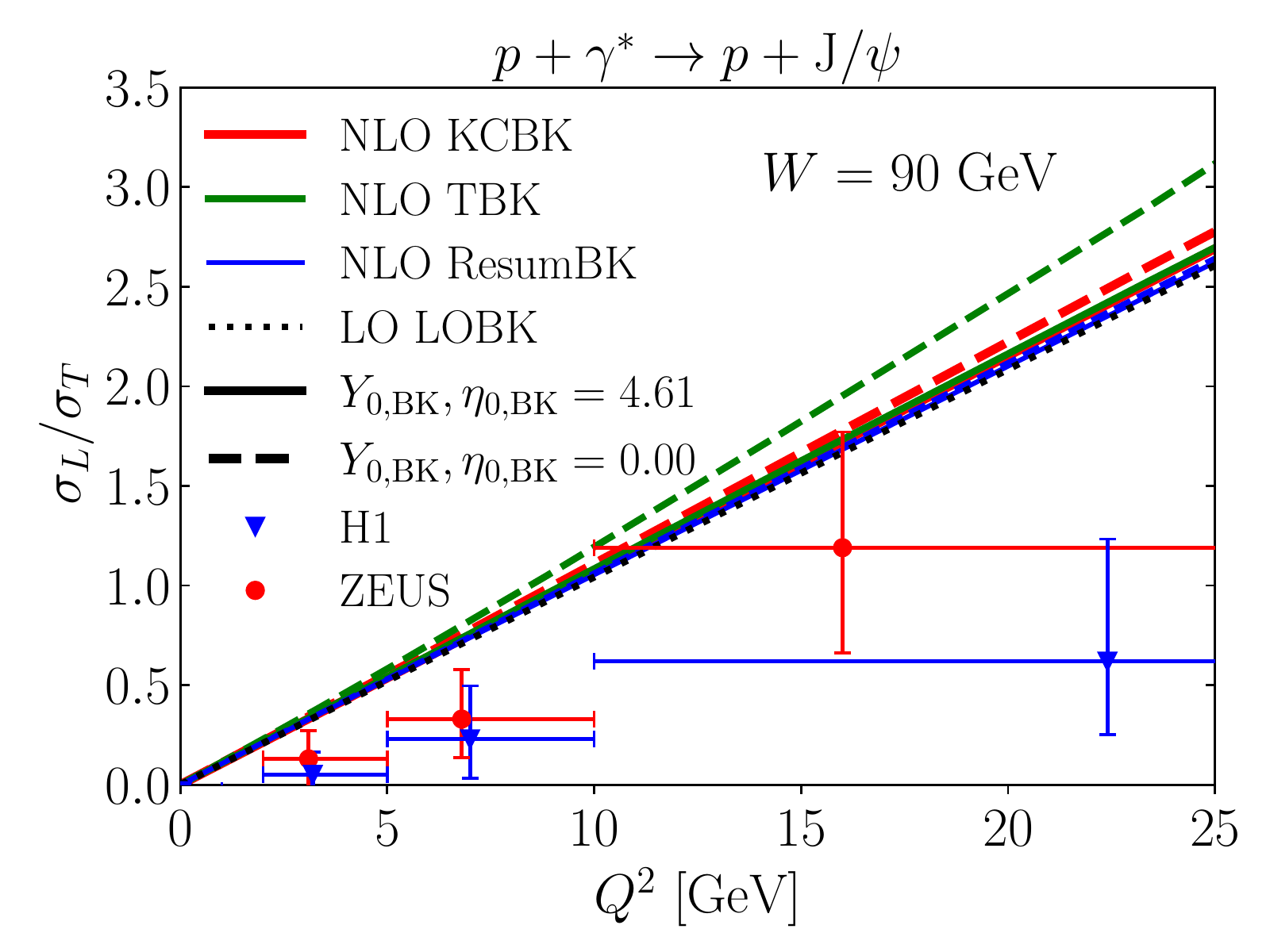}
        \caption{ Ratio in the nonrelativistic limit.
        }
        \label{fig:nonrel-R-W90}
    \end{subfigure}
    \begin{subfigure}{0.45\textwidth}
        \centering
        \includegraphics[width=\textwidth]{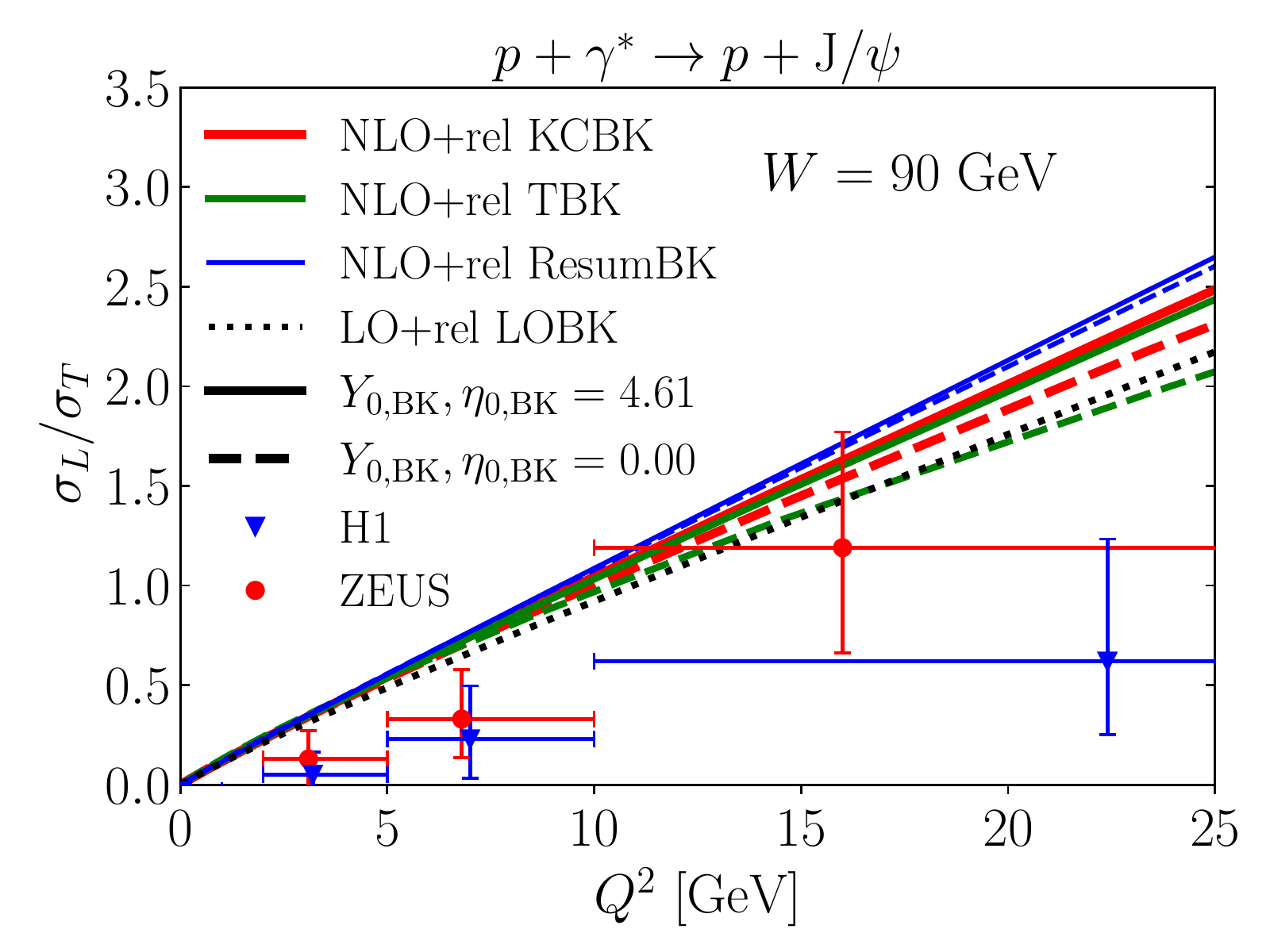}
        \caption{ Ratio with relativistic corrections. }
        \label{fig:rel-R-W90}
    \end{subfigure}
	   \caption{Longitudinal \jpsi production cross section divided by the transverse cross section compared to H1~\cite{H1:2005dtp} and ZEUS~\cite{ZEUS:2004yeh} data at $W=90\,\gev$. }
        \label{fig:R}
\end{figure}

Next we study the longitudinal-to-transverse \jpsi production cross section ratio where e.g. the normalization uncertainty cancels. This is plotted in Fig.~\ref{fig:R} where we show the results for the nonrelativistic case (Fig.~\ref{fig:nonrel-R-W90}) and with the relativistic corrections (Fig.~\ref{fig:rel-R-W90}), compared to the HERA data \cite{ZEUS:2004yeh, H1:2005dtp}. Excluding the TBK $\etabk=0.00$ dipole, the NLO corrections have only a modest effect on this ratio, slightly increasing it in general. In the nonrelativistic limit the differences between the different dipole amplitude fits having initial evolution rapidity $\Ybk=4.61$ (or $\etabk=4.61$) are negligible, whereas with $\Ybk=0$ ($\etabk=0$) there is some variation. Including the relativistic corrections increases this variation in both cases. 
The agreement with the experimental data is similar for both the LO and NLO results, and in both cases the ratio seems to be somewhat overestimated. 
This ratio is sensitive to the form of the wave function (see e.g. Ref~\cite{Lappi:2020ufv}), and in particular including the relativistic corrections improves the agreement with the HERA data slightly.

\begin{figure}
	\centering
    \includegraphics[width=0.6\textwidth]{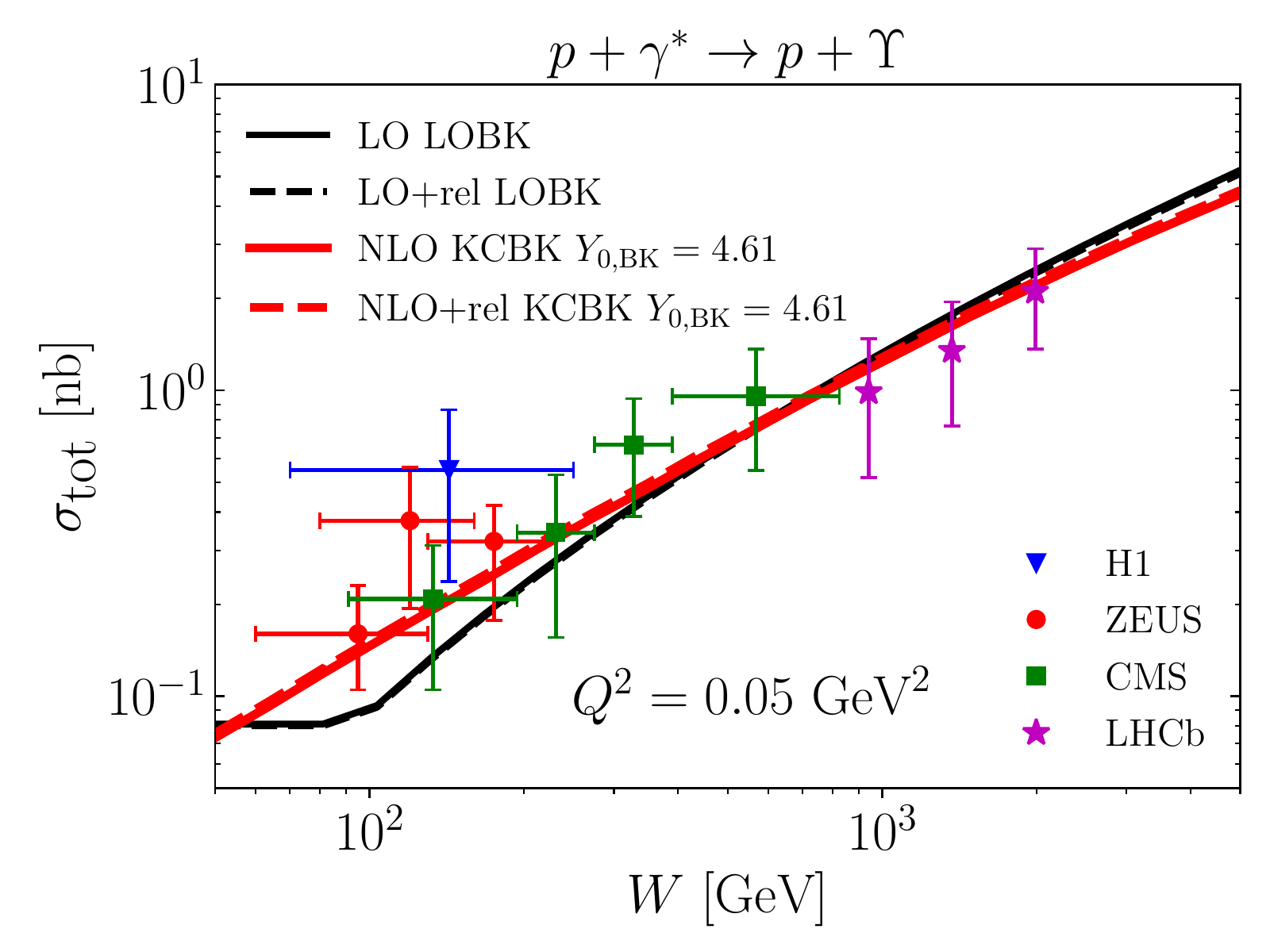}
	 \caption{Exclusive $\Upsilon$ photoproduction cross section as a function of center-of-mass energy $W$ compared with HERA~\cite{H1:2000kis, ZEUS:1998cdr, ZEUS:2009asc} and LHC~\cite{CMS:2018bbk, LHCb:2015wlx} data. Results are obtained using the KCBK evolution equation with initial evolution rapidity $\Ybk=4.61$.
	 }
        \label{fig:xs_upsilon}
\end{figure}

Finally, we consider exclusive $\Upsilon$ photoproduction. As $\Upsilon$ is much heavier than the \jpsi studied above, relativistic corrections become very small and it can be expected to be more sensitive to the next-to-leading order QCD corrections. The $\Upsilon$ photoproduction cross section as a function of the center-of-mass energy $W$ is shown in Fig.~\ref{fig:xs_upsilon} and compared with HERA~\cite{H1:2000kis, ZEUS:1998cdr, ZEUS:2009asc}, CMS~\cite{CMS:2018bbk} and LHCb~\cite{LHCb:2015wlx} data.
The $t$-integration of the analytic result is done using Eq.~\eqref{eq:xs_t-dependence} with the experimentally measured $W$-independent slope parameter $b=4.3 \gev^{-2}$~\cite{ZEUS:2011spj}.
The relativistic corrections, calculated using the NRQCD matrix elements from Ref.~\cite{Chung:2010vz}, are indeed small at $\sim 3\%$  level at all $W$. The next-to-leading order contributions are larger, and result in a slower $W$ dependence compared to leading-order results. Results obtained at leading and next-to-leading order are both compatible with the available data. Again the leading-order result is constant at $W<95\gev$ where $\xpom > 0.01$.

\section{Conclusions}
\label{sec:conclusions}

We have presented the first calculation for transversely polarized exclusive heavy vector meson production at next-to-leading order accuracy in the Color Glass Condensate framework. The main result of this work is Eq.~\eqref{eq:total_NLO_wf}, which is the scattering amplitude for the transverse vector meson production at NLO in the nonrelativistic limit. We have also presented how relativistic corrections, which are generally as important numerically as the NLO QCD corrections (in the case of \jpsi production), can be consistently included in the NLO calculation. The corresponding part of the scattering amplitude is given in Eq.~\eqref{eq:rel_corrections_NRQCD}.
Combined with the NLO calculation for the longitudinal production presented in Ref.~\cite{Mantysaari:2021ryb}, the results of this paper allow for phenomenological studies of heavy vector meson production at next-to-leading order accuracy. 

The NLO corrections are numerically significant for both the transverse and longitudinal production amplitude. This is largely compensated by the smaller dipole amplitude in the NLO calculation,
making the NLO results mostly in line with the LO production and rendering the NLO corrections generally moderate. When the first relativistic corrections are added the agreement of the coherent \jpsi production cross section with the HERA and LHC data is improved, especially at small photon virtualities where the relativistic corrections are larger than the NLO corrections. 

If the NLO cross sections are calculated using dipole amplitude fits from Ref.~\cite{Beuf:2020dxl} where there is a long evolution before one enters the region constrained by the small-$x$ structure function data ($\Ybk=0$ or $\etabk=0$ fits), the NLO corrections become very large at small center-of-mass energies and even result in negative cross sections. However, we also note that the nonperturbative parameters describing the initial condition in these fits are not physically well motivated. Large NLO corrections observed in this  case illustrate how heavy particle production is sensitive to different length scales than structure function calculations and can provide additional constraints when the nonperturbative initial condition for the Balitsky-Kovchegov equation is determined.

Now that the results for both longitudinal and transverse~\cite{Beuf:2021qqa,Beuf:2021srj,Beuf:2022ndu} photon wave functions with massive quarks are available, it will be possible to extend the dipole amplitude fits of Ref.~\cite{Beuf:2020dxl} to the massive quark case. This will allow for a consistent phenomenological study of NLO vector meson production at $t=0$, which was not possible in this paper. 
Furthermore, as the impact parameter dependence of the gluon structure is directly related to the $t$ dependence of vector meson production, it would be especially interesting to study $t$-dependent vector meson production amplitudes. This requires additional modeling for the impact parameter dependence of the dipole amplitude, which is the reason it was not considered in this work.

With these possible future developments in mind, the results presented in this paper can be used for extensive comparisons with heavy vector meson production data from HERA~\cite{H1:2013okq,ZEUS:2002wfj,ZEUS:2004yeh} and from the UPC physics program at the LHC~\cite{LHCb:2014acg,ALICE:2018oyo,LHCb:2014acg, LHCb:2018rcm}, along with making predictions for the future EIC. The results can also be extended from  proton targets to heavy nuclei by changing the dipole-target scattering amplitude, which enables studies of non-linear QCD dynamics in heavy nuclei at small-$x$. This is especially interesting given the existing and future data from ultra-peripheral Pb+Pb collisions at the LHC~\cite{ALICE:2012yye,ALICE:2013wjo,CMS:2016itn,ALICE:2021gpt,ALICE:2019tqa,LHCb:2021bfl} and possibilities at the future nuclear DIS experiments. 

\section*{Acknowledgements}

We thank M. Escobedo, T. Lappi and R. Paatelainen for useful discussions and are grateful to authors of Refs.~\cite{Beuf:2021srj,Beuf:2022ndu} for sharing their results before publication.
This work was supported by the Academy of Finland, the Centre of Excellence in Quark Matter, and projects 338263, 346567 (H.M), and 321840 (J.P), by the Finnish Cultural Foundation (J.P), and under the European Union’s Horizon 2020 research and innovation programme by the European Research Council (ERC, grant agreement No. ERC-2018-ADG-835105 YoctoLHC) and by the STRONG-2020 project (grant agreement No. 824093). The content of this article does not reflect the official opinion of the European Union and responsibility for the information and views expressed therein lies entirely with the authors.

\bibliographystyle{JHEP-2modlong.bst}
\bibliography{refs}

\clearpage

\appendix

\section{Dependence on the wave function renormalization scheme}
\label{appendix:scheme-dep}
As discussed in Sec.~\ref{sec:relativistic}, the renormalization of the LOWF can be done in different ways. In this paper we consider two different renormalization schemes, called the \emph{decay width} scheme (Eq.~\eqref{eq:LOWF_decay_width_scheme}) and the \emph{wave function} scheme (Eq.~\eqref{eq:LOWF_wave_function_scheme}). 
The reason for the different schemes is that the decay width scheme is convenient in the nonrelativistic case as then one can use the same running coupling constant \eqref{eq:running_coupling} as in the rest of the calculation when renormalizing the meson wave function. On the other hand the wave function scheme is necessary when considering the relativistic corrections as in that case the decay width scheme is not possible for transverse production.
This makes the NLO cross section dependent on the choice of the wave function renormalization scheme. The choice of the scheme appears parametrically at $\as^2$, and is thus of higher order than we consider here, but it can still have an effect on the numerical results. This is what we will study in this Appendix.

In the nonrelativistic case we can choose to use either of these two schemes. The decay width scheme is used as described by Eq.~\eqref{eq:LOWF_decay_width_scheme}. For the wave function scheme, we can calculate the dimensionally regularized LOWF in Eq.~\eqref{eq:LOWF_wave_function_scheme} from the nonrelativistic limit of the leptonic width using Eq.~\eqref{eq:dim_reg_LOWF}.
In that case one has to choose the scale at which to calculate the coupling constant. A natural choice is the mass of the vector meson which in the case of \jpsi evaluates to $\as(M_{\jpsim}) \approx 0.25$ \cite{Bodwin:2007fz} (for $\Upsilon$ this is $\as(M_\Upsilon) \approx 0.18$ \cite{Chung:2010vz}).
The difference between the decay width and wave function schemes is then where the  NLO contribution to the decay width appears when calculating the NLO production amplitude. In the decay width scheme it is calculated as part of the virtual correction $\kcal^\nlo_{q\bar q}$; in the wave function scheme it appears when we calculate the value of the dimensionally regularized LOWF. 

\begin{figure}
	\centering
    \begin{subfigure}{0.45\textwidth}
        \centering
        \includegraphics[width=\textwidth]{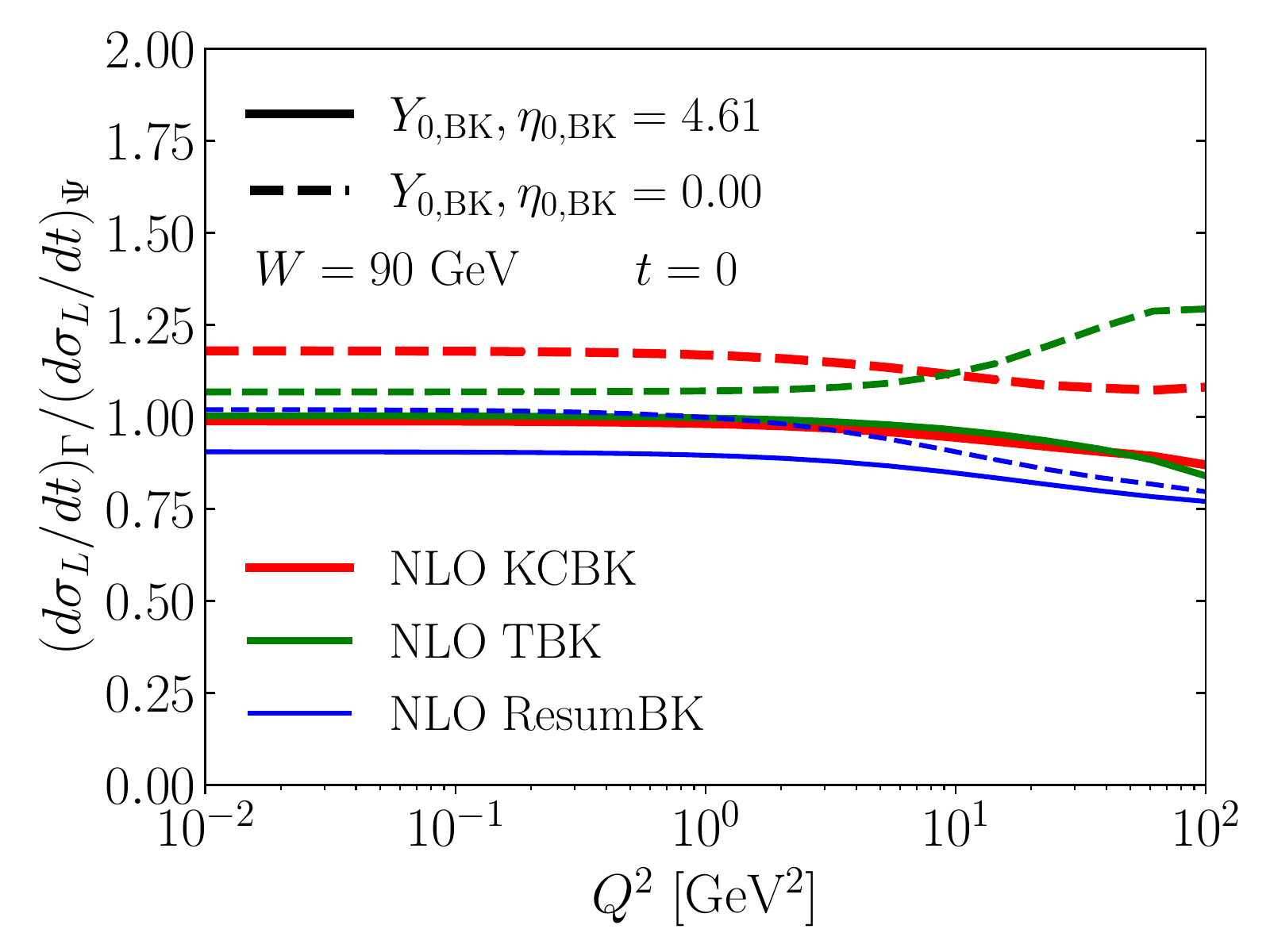}
        \caption{ Longitudinal production as a function of virtuality. }
        \label{fig:scheme-dep-L-W_90}
    \end{subfigure}
    \begin{subfigure}{0.45\textwidth}
        \centering
        \includegraphics[width=\textwidth]{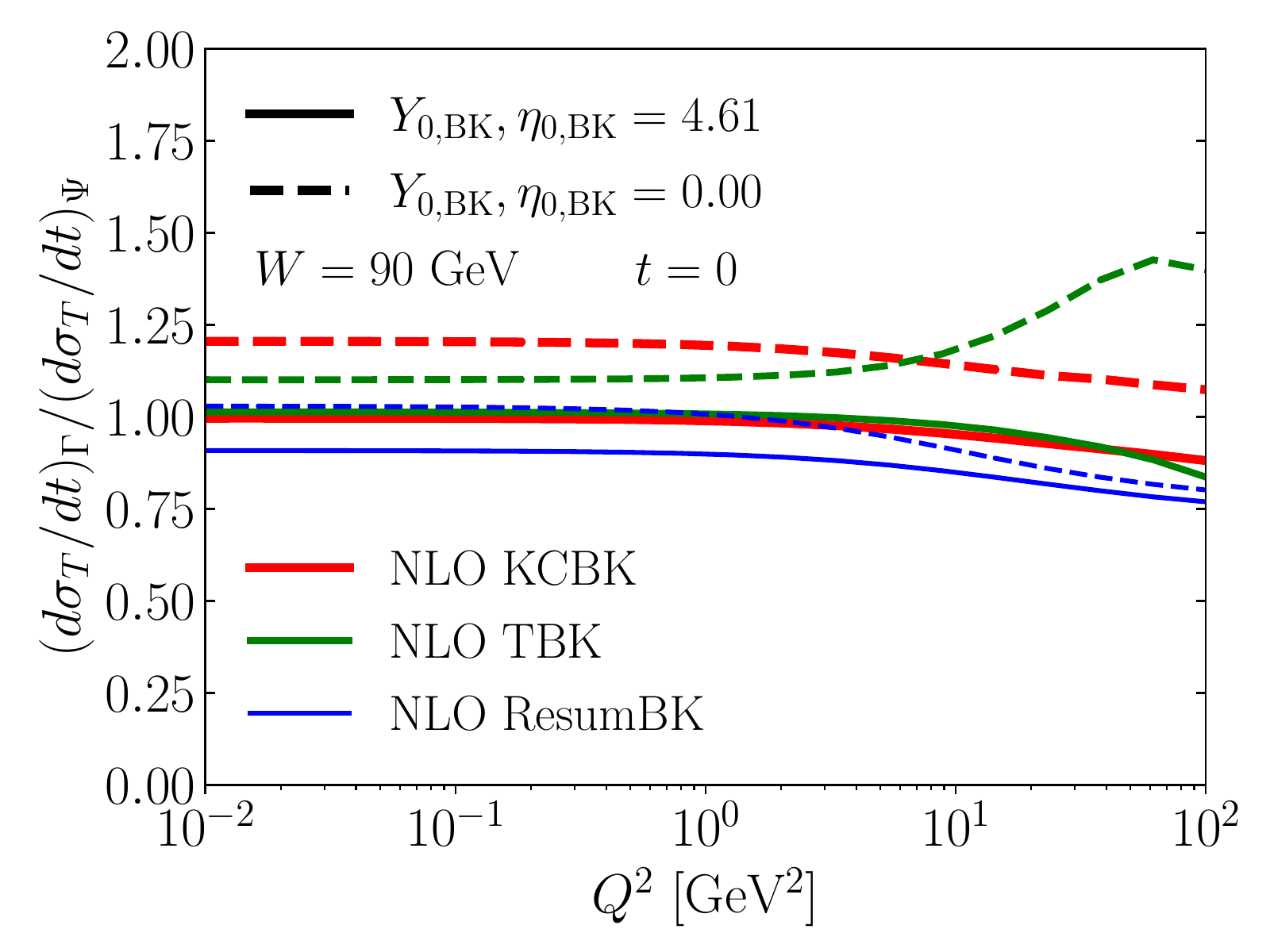}
        \caption{ Transverse production as a function of virtuality.}
        \label{fig:scheme-dep-T-W_90}
    \end{subfigure}
    \begin{subfigure}{0.45\textwidth}
        \centering
        \includegraphics[width=\textwidth]{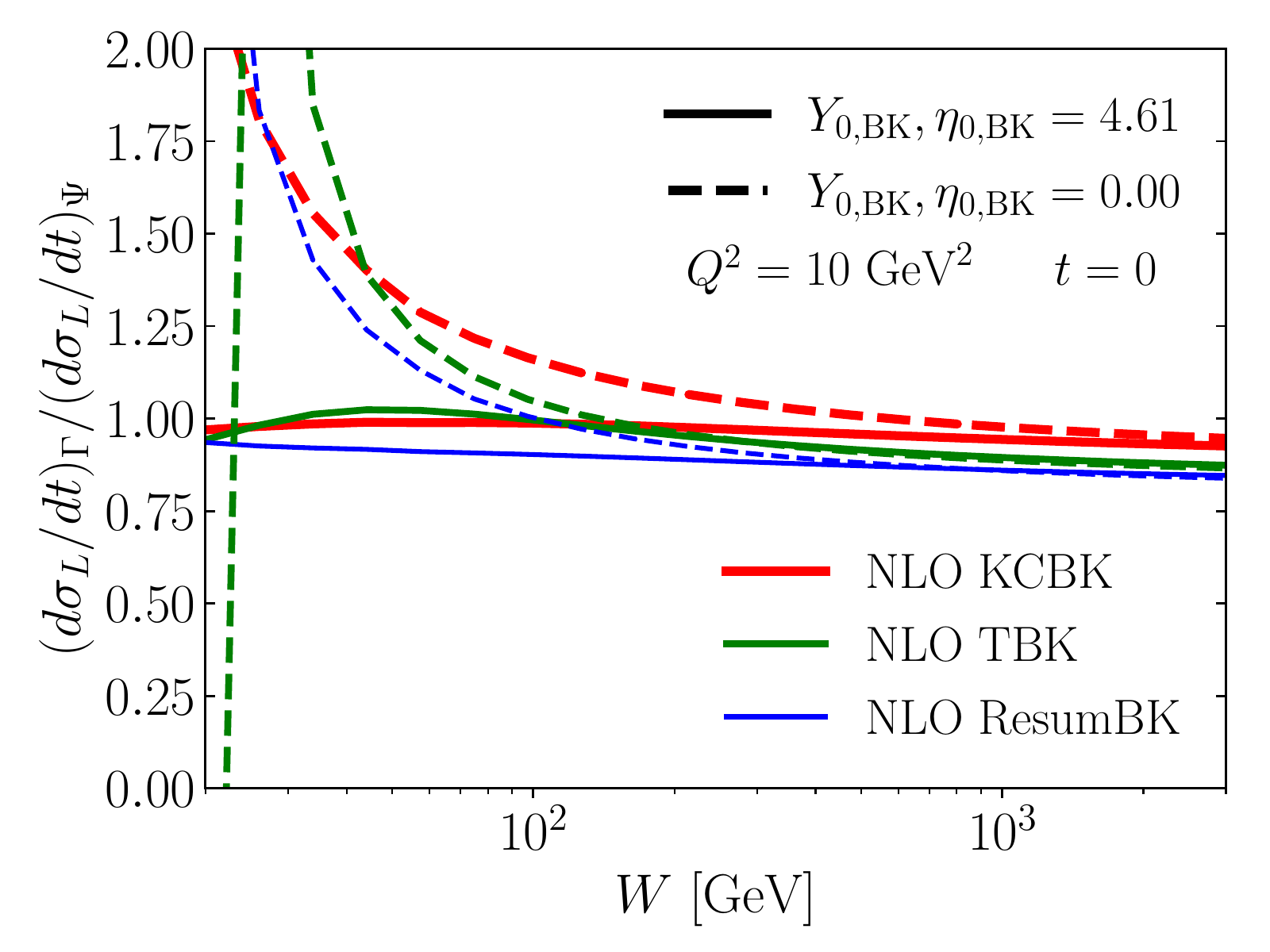}
        \caption{ Longitudinal production as a function of center-of-mass energy.  }
        \label{fig:scheme-dep-L-Q2_10}
    \end{subfigure}
    \begin{subfigure}{0.45\textwidth}
        \centering
        \includegraphics[width=\textwidth]{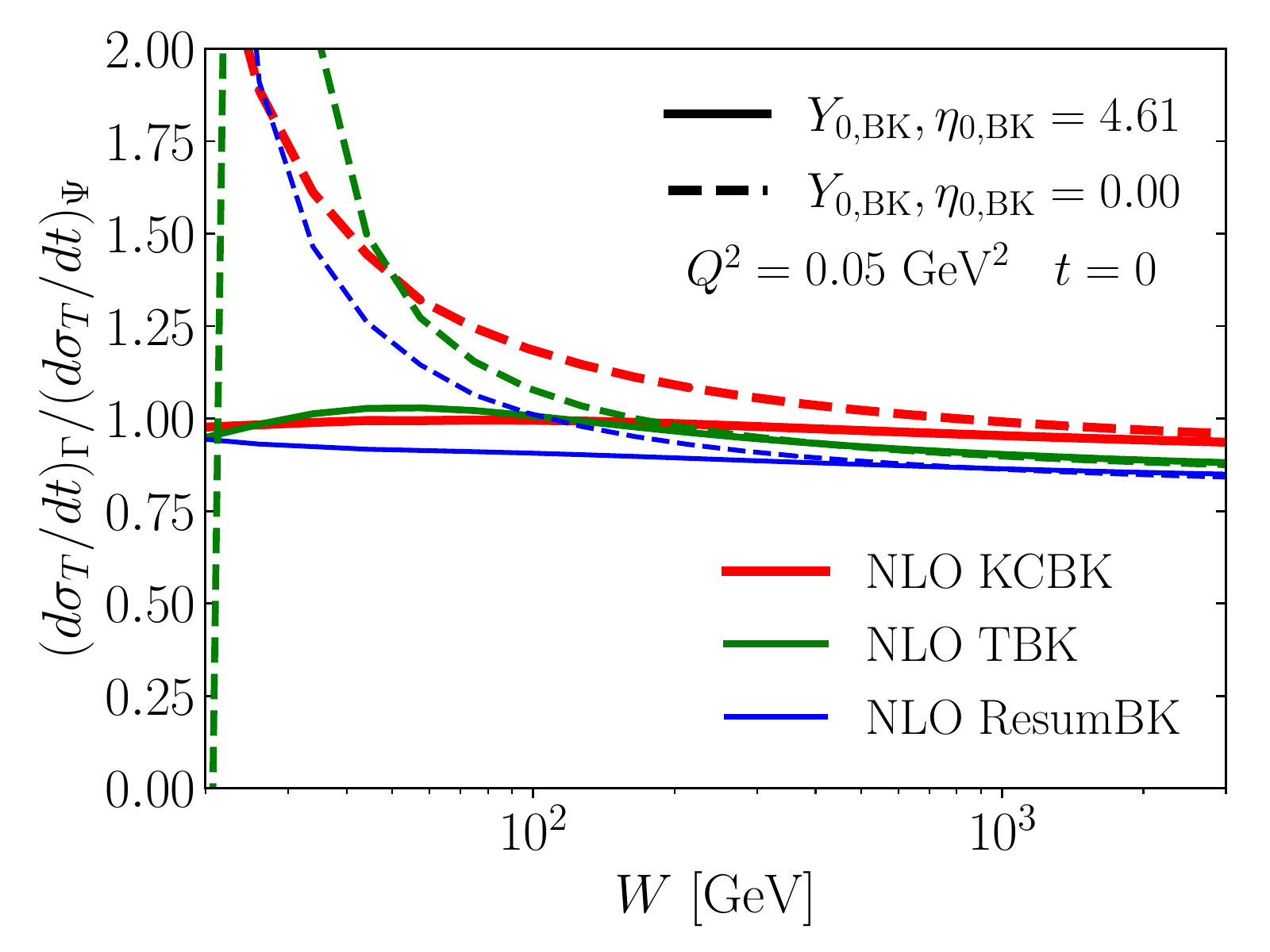}
        \caption{Transverse production as a function of center-of-mass energy.  }
        \label{fig:scheme-dep-T-Q2_0.05}
    \end{subfigure}
	   \caption{Scheme dependence of the differential \jpsi production cross section at NLO illustrated as a ratio of the cross sections calculated in the decay width ($\Gamma$) and wave function ($\Psi$) schemes in the nonrelativistic limit. }
        \label{fig:scheme_dep}
\end{figure}

To quantify the effects of the scheme choice, we have evaluated the NLO differential cross section for exclusive \jpsi production as a function of $Q^2$ and $W$ in the nonrelativistic case using the two different schemes. The ratio of these cross sections is shown in Fig.~\ref{fig:scheme_dep} for both longitudinal and transverse production. We see that in the  calculations where a  BK evolution starting at rapidity $\Ybk = 4.61$ (or $\etabk=4.61$ in the case of TBK evolution) is used there is only a small dependence on the scheme, of the order $10\%$. On the other hand, if the BK evolution starts at initial rapidity $0$ the differences between the two schemes become very large at small center-of-mass energies $W$. However, we note that as discussed in Sec.~\ref{sec:numerical} the initial conditions for evolutions starting at rapidity $\Ybk=0$ have unphysical features that do not strongly affect the structure function calculations and the fit process of Ref.~\cite{Beuf:2020dxl}, but have a large effect here as heavy vector meson production is sensitive to smaller dipole sizes.

\section{Longitudinal vector meson production at next-to-leading order}
\label{appendix:longitudinal}
For completeness, we list here the expressions for longitudinal vector meson production at NLO from Ref.~\cite{Mantysaari:2021ryb}. The NLO production amplitude can be divided into similar parts as in the case of transverse production. In the decay width scheme, the production amplitude can be written as
\begin{multline}
	\label{eq:total_NLO_longitudinal}
		-i \acal^L = -\sqrt{ \Gamma(V \rightarrow e^- e^+) \frac{3 \pi  M_V}{2N_c e_f^2 e^4} } \sqrt{\frac{N_c}{2}} \frac{e e_f Q}{2 \pi} 2 \int \dd[2]{\xt_{01}} \int \dd[2]{\bt} 
		\Bigg\{ \kcal_{q \bar q}^\lo(Y_0) \\
		+\frac{\as C_F}{2\pi} \kcal_{q \bar q, \Gamma}^{\nlo,L}(\Ydip) 
		+ \frac{\as C_F}{2\pi}\int \dd[2]{\xt_{20}} \int_{\zmin}^{1/2} \dd[]{z_2} \kcal_{q\bar q g}^L(Y_\text{qqg})\Bigg\}
\end{multline} 
where $\kcal^\lo_{q \bar q}$ is defined in Eq.~\eqref{eq:K_qq_LO},
\begin{multline}
    \label{eq:K_qq_NLO_longitudinal}
        \kcal_{q \bar q, \Gamma}^{\nlo,L}(\Ydip) = 
         N_{01}(\Ydip)\times\Bigg[
        \mathcal{\tilde I}_\nu\left(\frac{1}{2},\xt_{01}\right)+\kcal^L
        \\
        +K_0(\zeta) \left(\Omega_\vcal^L\left(\gamma;\frac{1}{2}\right) + L\left(\gamma;\frac{1}{2}\right)-\frac{\pi^2}{3}+2+4-3\ln(\frac{|\xt_{01}|m_q}{2})-3\gamma_E \right) \Bigg]
\end{multline} 
and
\begin{multline}
    \label{eq:K_qqg_longitudinal}
        \kcal_{q \bar q g}^L(\Yqqg) = -32 \pi m_q  \Bigg\{ \frac{i \xt_{20}^i}{|\xt_{20}|} K_1(2m_q z_2 |\xt_{20}|) \\
        \times\left[\left( (1-z_2)^2 + z_2^2 \right) \ical_{(j)}^i + (2z_2^2-1)(1-2z_2) \ical_{(k)}^i \right] N_{012}(\Yqqg) \\
		+4m_q z_2^3 K_0(2m_q z_2|\xt_{20}|) \left[ \ical_{(j)} - \frac{1-2z_2}{1+2z_2} \ical_{(k)} \right] N_{012}(\Yqqg) \\
		+\frac{1}{8\pi^2} \left((1-z_2)^2+z_2^2\right) \frac{1}{m_q z_2 |\xt_{20}|^2} K_0(\zeta) e^{-\xt_{20}^2\B} N_{01}(\Yqqg)
		\Bigg\}.
\end{multline}
The special functions $L(\gamma, z)$, $\ical_{(j)}$, $\ical_{(k)}$, $\ical^i_{(j)}$ and $\ical^i_{(k)}$ are defined in Sec.~\ref{sec:photon_wf}. The terms $\kcal^L$ and $\Omega_\vcal^L$ can be written as
\begin{multline}
	\label{eq:K_L}
		\kcal^L = \int_0^{1/2} \dd[]{z} \Bigg\{ 16z(1-z)K_0\left(|\xt_{01}|\sqrt{\overline Q^2+m_q^2}\right)2z \left[K_0(\tau)-\tau K_1(\tau) \right] \\
		+ \frac{1}{(z-1/2)^2}\Bigg\{ 
			16z(1-z)K_0\left(|\xt_{01}|\sqrt{ \overline Q^2+m_q^2}\right) \left[
				2z(1-z)K_0(\tau)-z\left(
					K_0(\tau)-\frac{\tau}{2}K_1(\tau)
				\right)
			\right] \\
			- K_0(\zeta) \left[
				1+2\left(z-\frac{1}{2}\right) \left[
					1+2 \gamma_E + 2\ln(m_q|\xt_{01}|)+2\ln\left(\frac{1}{2}-z\right)
				 \right]
			\right]
		\Bigg\}\Bigg\}
\end{multline}
and
\begin{multline}
	\label{eq:Omega_L}
	\Omega_\vcal^L(\gamma;z) = \frac{1}{2z} \left[ \ln(1-z) + \gamma \ln(\frac{1+\gamma}{1+\gamma-2z}) \right]\\
	+\frac{1}{2(1-z)} \left[ \ln(z) + \gamma \ln(\frac{1+\gamma}{1+\gamma-2(1-z)}) \right] \\
	+\frac{1}{4z(1-z)} (\gamma-1) \ln(\frac{\overline Q^2+m_q^2}{m_q^2}) +\frac{m_q^2}{2\overline Q^2} \ln(\frac{\overline Q^2+m_q^2}{m_q^2}).
\end{multline}
Finally, the special function $\mathcal{\tilde I}_\vcal(z,\xt_{01})$ is given by
\begin{equation}
\label{eq:Iv}
    \mathcal{\tilde I}_\vcal(z,\xt_{01}) =\mathcal{\tilde I}_{\vcal_{(a)+(b)}} (z,\xt_{01}) + \mathcal{\tilde I}_{\vcal_{(c)+(d)}}(z,\xt_{01}),
    \end{equation} 
    with 
\begin{multline}
	\label{eq:J1_L}
	\mathcal{\tilde I}_{\vcal_{(a)+(b)}}(z,\xt_{01})  = \int_0^1 \frac{\dd[]{\xi}}{\xi} \left(- \frac{2\ln \xi}{1-\xi}+\frac{1+\xi}{2} \right) \\
	\times\left[ 2K_0 \left(|\xt_{01}| \sqrt{\overline Q^2+m_q^2}\right) 
	 -K_0 \left(|\xt_{01}| \sqrt{\overline Q^2+m_q^2+\frac{(1-z)\xi}{1-\xi}m_q^2}\right)
	  \right. \\
	  - \left. K_0 \left(|\xt_{01}| \sqrt{\overline Q^2+m_q^2+\frac{z\xi}{1-\xi}m_q^2}\right)
		\right]
\end{multline}
and
\begin{multline}
	\label{eq:J2_L}
	\mathcal{\tilde I}_{\vcal_{(c)+(d)}}(z,\xt_{01}) = m_q^2 \int_0^1 \dd{\xi} \int_0^1 \dd{x} \\
	 \Bigg\{ \left[ K_0 \left(|\xt_{01}| \sqrt{\overline Q^2+m_q^2}\right)-K_0 \left(|\xt_{01}| \sqrt{\frac{\overline{Q}^2+m_q^2}{1-x}+\kappa}\right)\right]
	\\
	\times\frac{C^L_m}{(1-\xi)(1-x)\left[x(1-\xi)+\frac{\xi}{1-z}\right]\left[\frac{x(\overline Q^2+m_q^2)}{1-x}+\kappa\right]}   \\
	+ (z \to 1-z) \Bigg\}.
\end{multline}
The coefficient $C_m^L$ in the above expression reads
\begin{align}
	\label{eq:Clm}
	C^L_m &= \frac{z^2 (1-\xi)}{1-z} \left[ -\xi^2 +x(1-\xi) \frac{1+(1-\xi) \left(1+\frac{z \xi}{1-z}\right)}{x(1-\xi) + \frac{\xi}{1-z}} \right],
\end{align}
and %
\begin{equation}
	\label{eq:kappa}
 	\kappa = \frac{\xi m_q^2}{(1-\xi)(1-x)\left[x(1-\xi)+\frac{\xi}{1-z}\right]} \left[\xi(1-x)+x \left(1-\frac{z(1-\xi)}{1-z}\right) \right].
\end{equation}

When considering relativistic corrections, one can no longer use the decay width scheme for transverse production. It is then more consistent to use the wave function scheme which works also with the relativistic corrections. In the wave function scheme, the longitudinal production amplitude can be written as
\begin{multline}
	\label{eq:total_NLO_wf_longitudinal}
		-i \acal^L = -  \int \frac{\dd[]{z'}}{4 \pi} \phi^{q \bar q,L}_\textrm{DR} \times
		\sqrt{\frac{N_c}{2}}\frac{e e_f Q}{2\pi } 2\int \dd[2]{\xt_{01}}\int \dd[2]{\bt}
		  \Bigg\{ \kcal_{q \bar q}^\lo(Y_0) \\
		+\frac{\as C_F}{2\pi}  \kcal_{q \bar q, \Psi}^{\nlo,L}(\Ydip) 
		+ \frac{\as C_F}{2\pi}\int \dd[2]{\xt_{20}} \int_{\zmin}^{1/2} \dd[]{z_2} \kcal_{q\bar q g}^L(\Yqqg)\Bigg\}
\end{multline}
where the LOWF $\phi^{q \bar q,L}_\textrm{DR} $ is now the dimensionally regularized one from Eq.~\eqref{eq:LOWF_wave_function_scheme}, and the virtual correction becomes
\begin{multline}
	\kcal_{q \bar q, \Psi}^{\nlo,L}(\Ydip) =  N_{01}(\Ydip) \times \left[ 
        \mathcal{\tilde I}_\nu\left(\frac{1}{2},\xt_{01}\right)+\kcal^L
        \right.\\
        \left.+K_0(\zeta) \left(\Omega^L_\vcal\left(\gamma;\frac{1}{2}\right) + L\left(\gamma;\frac{1}{2}\right)-\frac{\pi^2}{3}+2-3\ln(\frac{|\xt_{01}|m}{2})-3\gamma_E \right) \right]
\end{multline}
instead of $\kcal_{q \bar q, \Gamma}^\nlo(\Ydip)$. The relativistic corrections to longitudinal production can be written as
\begin{equation}
	\label{eq:rel_longitudinal}
	\begin{split}
		-i \acal^L_\text{rel} =& -\frac{e e_f Q \sqrt{N_c}}{2\pi}2 \int \dd[2]{\xt_{01}} N_{01}(\Ydip) \\
		&\times\frac{1}{4} \Bigg\{ \left( \phi^{q \bar q,L}_{+-}(0,0,2)+\phi^{q \bar q,L}_{-+}(0,0,2) \right)  \left[ 2K_0(\zeta)-\frac{Q^2 \xt_{01}^2}{4\zeta}K_1(\zeta) \right]\\
		&+\left(\phi^{q \bar q,L}_{+-}(2,0,0)+\phi^{q \bar q,L}_{-+}(2,0,0)+\phi^{q \bar q,L}_{+-}(0,2,0)+\phi^{q \bar q,L}_{-+}(0,2,0)\right)  \frac{m^2 \xt_{01}^2}{2} K_0(\zeta) \Bigg\} \\
		=&-\frac{e e_f Q \sqrt{N_c}}{2\pi}2 \int \dd[2]{\xt_{01}} N_{01}(\Ydip) \\
		&\times \frac{1}{2}\Bigg\{  \phi^{q \bar q,L}_{+-}(0,0,2) \left[ 2K_0(\zeta)-\frac{Q^2 r^2}{4\zeta}K_1(\zeta) \right]
		+\phi^{q \bar q,L}_{+-}(2,0,0)  m^2 \xt_{01}^2 K_0(\zeta) \Bigg\}.
	\end{split}
\end{equation}
where the expression was simplified using the identity $\phi^{q \bar q,L}_{+-}(\rt,z') =\phi^{q \bar q,L}_{-+}(\rt,z')$  that follows from the spin-parity of the vector meson as discussed in Sec.~\ref{sec:relativistic}. 

The nonperturbative constants related to the LOWF can be written in terms of the rest-frame wave function $\phi_\textrm{RF}(\vec r)$ \cite{Lappi:2020ufv}, giving us
\begin{equation}
     \int \frac{\dd[]{z'}}{4 \pi} \phi^{q \bar q, L}_\textrm{DR}  =\sqrt{2} \phi^{q \bar q,L}_{+-}(0,0,0)= \frac{1}{\sqrt{4 m_q}} \left[\phi_\text{RF}(0) + \frac{5}{12m_q^2}\vec \nabla^2 \phi_\text{RF}(0)\right]
\end{equation}
\begin{equation}
    \phi^{q \bar q,L}_{+-}(2,0,0) = \phi^{q \bar q,L}_{-+}(0,0,2) = \frac{1}{6\sqrt{2m_q}} \frac{1}{m_q^2}\vec \nabla^2 \phi_\text{RF}(0).
\end{equation}
This allows us to write the $v^2$ relativistic correction in the compact form
\begin{multline}
	\label{eq:rel_NRQCD_longitudinal}
		-i \acal^L_\text{rel} = -\frac{e e_f Q \sqrt{N_c}}{2\pi\sqrt{2}}2 \int \dd[2]{\xt_{01}}  \int\dd[2]{\bt} N_{01}(\Ydip) \\
		\times \frac{\nabla^2 \phi_\text{RF}(0)}{12m_q^2\sqrt{m_q}} 
		 \left[ 2K_0(\zeta)-\frac{Q^2 \xt_{01}^2}{4\zeta}K_1(\zeta)+ m_q^2 \xt_{01}^2 K_0(\zeta) \right].
\end{multline}
The value of the rest-frame wave function and its derivatives can be related to NRQCD matrix elements as described in Sec.~\ref{sec:relativistic}.

\end{document}